\documentclass[aps, prd,superscriptaddress, keywords,pacs,nofootinbib,amsmath,amssymb,floatfix]{revtex4-2}
\usepackage{graphicx}
\usepackage{dcolumn}
\usepackage{booktabs,array,tabularx,longtable}
\usepackage{bm}
\usepackage{hyperref}
\usepackage{xcolor}
\usepackage[margin=1.5cm]{geometry}
\usepackage{tikz}
\usepackage[compat=1.1.0]{tikz-feynman}
\usetikzlibrary{arrows.meta}
\usetikzlibrary{arrows.meta,positioning,calc}
\usepackage{adjustbox}
\tikzfeynmanset{warn luatex=false}
\tikzfeynmanset{warn luatex=false}

\usepackage{tabularx}
\usepackage{array}
\usepackage{ragged2e}

\usepackage{pgfplots}
\usepgfplotslibrary{groupplots}
\pgfplotsset{compat=1.18}

\definecolor{protonblue}{RGB}{31,78,121}
\definecolor{waterblue}{RGB}{214,236,250}
\definecolor{goldNP}{RGB}{230,170,20}
\definecolor{radicalred}{RGB}{190,45,45}
\definecolor{dnagreen}{RGB}{45,125,75}

\tikzset{
  protonline/.style={
    -{Latex[length=3mm,width=2mm]},
    very thick,
    draw=protonblue
  },
  reaction/.style={
    -{Latex[length=3mm,width=2mm]},
    thick
  },
  processbox/.style={
    rectangle,
    rounded corners=3pt,
    draw=black,
    line width=0.9pt,
    align=center,
    minimum height=14mm,
    minimum width=34mm,
    fill=white
  }
}

\begin{document}
\pagestyle{plain}

\title{A Multiscale Kinetic Model of DNA Damage Signalling and Repair under Combined Proton and Nanoparticle Exposure}
\author{Mushtaq Loan}
\affiliation{Department of Physics, Fairleigh Dickinson University, Teaneck, NJ 07666, United States}


\date{\today}

\begin{abstract}

\textbf{Background and Purpose:}
Gold nanoparticles (GNPs) can modify the spatial, physical, and
chemical characteristics of proton energy deposition, but an increase
in DNA-damage induction or signalling does not necessarily imply
efficient or correct repair. A mechanistic description must therefore
distinguish the initial lesion per nucleus from damage complexity, repair
kinetics, pathway allocation, and repair fidelity. We present a
multiscale kinetic framework for predicting the induction, processing,
and biological consequences of DNA double-strand breaks (DSBs) in
AG1522 human fibroblasts exposed to combined proton and 15-nm GNP
irradiation.

\textbf{Methods:}
The model connects proton dose and linear energy transfer (LET),
perinuclear GNP localization, secondary-electron and reactive-species
production, and Poisson--Gamma-distributed lesion induction. Initial
DSBs are separated into simple and complex classes and coupled to an
ATM--$\gamma$H2AX--53BP1 signalling cascade, cell-cycle-dependent
allocation among canonical non-homologous end joining, homologous
recombination, alternative end joining, and single-strand annealing,
and class-specific repair fidelities. The  post-irradiation
kinetics impose $k_{SC}=0$, thereby removing unsupported conversion of
simple DSBs into complex lesions. A positive, dose-independent
LET--GNP response surface was fitted for the complex-lesion repair
coefficient $k_C$. Model performance was assessed using
leave-one-LET-out and leave-one-GNP-concentration-out validation, in
which complete irradiation blocks were excluded before endpoint
prediction.

\textbf{Results:}
Increasing LET and GNP concentration shifted the initial lesion
spectrum toward complex DSBs and progressively reduced their repair
fidelity. The predicted simple-DSB fidelity remained
high, decreasing from $95.5\%$ to $91.2\%$, whereas complex-DSB
fidelity declined from $70.8\%$ at
$10~\mathrm{keV}\,\mu\mathrm{m}^{-1}$ and $5\times$ GNP concentration
to $8.8\%$ at $50~\mathrm{keV}\,\mu\mathrm{m}^{-1}$ and $20\times$.
The corresponding processed-lesion-weighted misrepair fraction
increased from $14.2\%$ to $70.0\%$. The $k_C$ values ranged
from $0.0828$ to $0.0292~\mathrm{h}^{-1}$, corresponding to
complex-DSB half-lives of approximately $8.4$--$23.8$ h.  The predicted residual
damage at $24$h post-irradiation agreed closely with the Geant4-DNA data
at low doses. Agreement was weaker for
correct repair and misrepair with systematic overprediction of correct
repair and underprediction of misrepair.

\textbf{Conclusions:}
The model identifies the initial complex-lesion fraction and
its slower resolution, rather than post-irradiation conversion of
simple DSBs, as the principal determinants of persistent damage under
high-LET, GNP-enhanced irradiation. The residual-DSB results provide
strong  validation of the kinetic component.
However, the opposing biases in correct repair and misrepair reveal an
inconsistency between the class-specific fidelities and the supplied
integrated-fidelity endpoint. These repair-fate predictions should
therefore be regarded as model-constrained estimates pending
independent measurements of correct-end joining, mutation,
chromosome aberrations, and clonogenic survival.\\


%
\end{abstract}

\maketitle

\clearpage

\begin{table*}[h]
\caption{Principal symbols used in the model.}
\label{tab:model-symbols}
\centering
\begin{tabular}{cccccccc}
\hline
Symbol & Meaning & Units \\ \hline
 &  & \\
$D_p,\dot D_p$ & Proton absorbed dose and proton dose rate & Gy; Gy\,h$^{-1}$ \\
$L(\mathbf r,E),L_D$ & Local proton LET and nuclear dose-averaged LET & keV\,$\mu$m$^{-1}$ \\
$\Phi_p(E,\mathbf r,t)$ & Differential proton fluence & particles\,m$^{-2}$\,energy$^{-1}$\,time$^{-1}$ \\
$\mathcal C_e,\mathcal C_{\rm cyt},\mathcal C_n$ & Extracellular, cytoplasmic, and  & mass\,volume$^{-1}$ or particles\,volume$^{-1}$\\
                                                                                & nuclear or perinuclear nanoparticle concentrations & \\
$C_n$ & Relative GNP concentration index  & dimensionless relative concentration \\
       & used in the supplied dataset & \\
$d_{\rm NP}$ & Gold-nanoparticle diameter & nm \\
$R,\Omega_R$ & Concentration of DNA-reactive species and  & concentration; \\
 & integrated reactive-species exposure & concentration$\times$time \\
$G_R,G_{\rm NP}$ & Water-radiolysis and nanoparticle-mediated  & species\,energy$^{-1}$ or \\
 &   reactive-species yields  &  model-specific yield units     \\
$Y_p,Y_{\rm NP},Y_{\rm syn}$ & Proton-only, nanoparticle, and  & DSB\,nucleus$^{-1}$\,Gy$^{-1}$, with concentration \\
    &  LET--concentration DSB-yield coefficients    &  and LET factors as appropriate   \\
$N_0$ & Expected initial number of  & DSB\,nucleus$^{-1}$ \\
& D SB lesions per cell nucleus & \\
$f_c$ & Complex or clustered DSB fraction& dimensionless \\
$S,S_0$ & Time-dependent and initial  & lesions\,nucleus$^{-1}$ \\
& simple-lesion populations & \\
$C,C_0$ & Time-dependent and initial  & lesions\,nucleus$^{-1}$\\
    & complex-lesion populations &    \\
$C_r,C_p$ & Rapidly processed and persistent  & lesions\,nucleus$^{-1}$ \\
    & complex-lesion populations   & \\
$\rho_C$ & Fraction of complex lesions assigned  & dimensionless \\
    &  to the persistent compartment    &   \\
$A,H,P$ & Normalized activated ATM, $\gamma$H2AX signal,  & dimensionless \\
    & and recruited end-recognition machinery    & \\
$p_N,p_H,p_A,p_{\rm SSA}$ & Allocation fractions for c-NHEJ, HR, alternative & dimensionless \\
 &   end joining,  and single-strand annealing  &    \\
$\mu_{js}(L,C_n)$ & Fidelity of repair pathway $j$  $s$ & dimensionless \\
     &  for lesion class & \\  
$\mu_C^{\rm G1},\mu_C^{\rm G2}, $ & Effective complex-lesion fidelities in G1, late-S/G2,  & dimensionless \\
      $ \mu_C^{\rm async}$  &     and an asynchronous cell population & \\ 
$g$ & Late-S/G2 pathway-mixture weight & dimensionless \\
$\lambda_1,\lambda_2$ & Fast and persistent lesion-removal & h$^{-1}$ \\
&   rate coefficients & \\
$\eta$ & Binary lesion-interaction coefficient & h$^{-1}$\,lesion$^{-1}$ \\
$\theta_0,\sigma$ & End-interaction strength and s& model-dependent rate; length \\
   &   patial interaction range     &   \\
$R_c,R_m,U$ & Correctly repaired, misrepaired, and r & lesions\,nucleus$^{-1}$ \\
    &   esidual lesion populations  &   \\
$\alpha_\Gamma,\beta_\Gamma$ & Gamma-distribution shape and  & dimensionless shape;  \\
    &   rate parameters for track-level lesion intensity &   inverse-intensity rate\\
$\kappa_c$ & Clustered-fraction correction to the  & dimensionless \\
    &       quadratic survival coefficient & \\  \hline
\end{tabular}
\end{table*}

\clearpage
\section{Introduction}
Proton therapy provides clinically advantageous dose localization. However, at the microscopic scale, proton energy deposition is highly heterogeneous. As protons lose energy, their LET increases, resulting in ionization patterns that are more spatially concentrated and chemically complex. Consequently, LET affects not only the total number of DNA double-strand breaks (DSBs)  but also their spatial clustering, end structure, associated base damage, repairability, and the probability of correct rejoining. Track-structure simulations across the clinically relevant proton-LET range predict increased DSB yield and lesion complexity with higher LET, whereas experimental measurements of repair-focus kinetics indicate that clustered, particle-induced lesions are generally resolved more slowly than predominantly isolated low-LET damage \cite{Zhu2020,Nakajima2020}. A biological description based solely on absorbed dose is therefore insufficient for predicting the biological consequences of proton irradiation.\\

Gold nanoparticles have been identified as local radiosensitizers that can modify proton-induced energy deposition and radiation chemistry. Due to their high atomic number, large surface-to-volume ratio, and capacity for intracellular accumulation, GNPs can enhance short-range secondary-electron emission, water radiolysis, reactive oxygen species production, and oxidative stress \cite{Sotiropoulos2017,Tabbakh2024,Cunningham2021,Fuster2026}. The magnitude of these effects depends on particle size, concentration, surface functionalization, intracellular trafficking, and proximity to sensitive cellular targets. Notably, 15-nm GNPs tend to accumulate in the perinuclear region, which may substantially modify the chemical environment surrounding nuclear DNA without necessarily entering the nuclear interior. A uniform macroscopic dose-enhancement factor cannot fully represent this heterogeneous geometry. For example, a GNP located far from chromatin may alter cellular redox conditions while contributing little direct nanometre-scale energy deposition to DNA. In contrast, a perinuclear GNP may generate locally dense secondary-electron and radical fields near the nuclear envelope.\\

An elevated initial DSB burden alone does not determine the subsequent biological response. Lesions with readily ligatable termini can be repaired rapidly and with high fidelity, whereas complex DSBs containing damaged termini, short DNA fragments, proximal base lesions, or locally clustered strand breaks may remain unresolved, undergo end resection, or be processed via error-prone pathways. Damage recognition by ATM, formation of $\gamma$H2AX domains, and recruitment of end-recognition factors  serve as indicators of damage sensing and signaling, rather than direct measures of successful repair. Robust or persistent signaling may accompany effective damage recognition, but may also indicate slowly processed or irreparable lesions. Therefore, a mechanistic model must distinguish among initial lesion induction, lesion complexity, damage recognition, repair pathway allocation, repair rate, and repair fidelity.
Several computational frameworks \cite{McMahon2016,McMahon2021,Ingram2019,Schuemann2019} describe complementary parts of this
multiscale problem. Track-structure platforms predict physical
and physicochemical energy-deposition events and initial DNA lesions,
whereas repair models describe lesion processing, pathway engagement,
misrepair, and chromosome-level outcomes. A remaining challenge is to
connect these scales without inferring unobserved kinetic transitions
directly from endpoint yields. In particular, the ratio of initial
simple and complex DSB yields does not define a post-irradiation
conversion coefficient between the two lesion classes. Once a lesion
has been classified according to its initial local structure, an
apparent increase in complexity during repair is more appropriately
represented through class-specific processing, persistent
subpopulations, pathway-dependent fidelity, or lesion interactions
than through an unsupported first-order conversion of simple DSBs into
complex DSBs.\\

The present work develops a kinetic framework for AG1522
human fibroblasts exposed to protons in the presence of 15-nm
perinuclear GNPs. Proton dose and LET are linked to spatially
heterogeneous GNP-mediated physical and chemical enhancement, including
secondary-electron production and DNA-reactive species. The resulting
initial DSB lesions per nucleus are divided into simple and complex lesion classes,
Damage recognition is represented through an ATM--$\gamma$H2AX--end-recognition signalling
sequence, followed by cell-cycle-dependent allocation among c-NHEJ,
HR, alt-EJ, and SSA. Class- and pathway-specific fidelity functions
then separate correctly repaired lesions from misrepaired lesions,
while the unresolved populations define the residual-damage endpoint.
 A principal feature of the model is the elimination of post-irradiation simple-to-complex conversion. The simple and complex lesion populations evolve according to their respective removal kinetics rather than through an assumed transfer between lesion classes.
Where necessary, the complex population can be further subdivided into rapidly processed and persistent components. This approach attributes persistent damage to the initially complex lesion spectrum and its slower processing, rather than to an artificial conversion of simple double-strand breaks (DSBs) following irradiation. 

The aims of this study are therefore: (i) to connect proton LET and perinuclear GNP concentration to initial DSB yield and complexity; (ii) to describe damage recognition and cell-cycle-dependent repair pathway allocation; (iii) to predict the time evolution of simple and
 complex DSBs without an unsupported \(S\!\rightarrow C\) conversion; (iv) to estimate correct repair, misrepair, residual damage, chromosome-aberration, and survival endpoints; and (v) to evaluate the predictive behavior of the model against the Geant4-DNA dataset. In order to minimize reliance on pointwise calibration, model assessment uses blocked leave-one-LET-out and leave-one-GNP-concentration-out procedures, in which complete irradiation-condition blocks are excluded before prediction. These tests constitute  validation across the supplied simulation domain; however, they do not replace external validation against independent biological measurements.\\

\section{Theoretical Framework}
\subsection{Proton fluence, Dose and LET}
The cell nucleus is treated as a bounded volume $V_n$ containing spatially characterized chromatin and nanoparticles. Proton transport determines the local LET $L(\mathbf r)$, 
while chemistry is coarse-grained into a radical field. DSBs are classified as simple ($S$) or complex/clustered ($C$). Complexity is operational: lesions requiring end processing, 
containing nearby base damage or strand breaks, or displaying slow focus resolution are assigned to $C$. The principal symbols used in the model are given in Table \ref{tab:model-symbols}, 
and the physical, chemical, signalling, repair, and biological scales are shown in Table \ref{tab:model-scales}. We impose the following assumptions in the model: (i) the nanoparticle 
concentration $C_{NP}$ is an intracellular mass or number concentration; it is not equated to the administered concentration; (ii) the nuclear localisation fraction 
$f_{\mathrm{nuc}}$ is retained because bulk cellular uptake does not determine DNA-scale enhancement; (iii) nanoparticles are not radiation; they modify proton 
transport, radiochemistry and cellular response, and (iv) enhancement coefficients are material-, size-, coating-, cell-line-, oxygenation- and beam-dependent.\\


\begin{table*}[h]
\caption{Physical, chemical, signalling, repair, and biological scales
represented in the model. The quoted intervals indicate
characteristic model times.}
\label{tab:model-scales}
\centering
\begin{tabular}{cccccccc}  \hline
Stage & Characteristic time & Principal state variables & Dominant processes \\ \hline
Physical transport & $10^{-15}$--$10^{-12}$ s & $\Phi_p(E,\mathbf r,t)$, $L(E,\mathbf r)$, $D_p(\mathbf r)$ & Proton transport, ionization, excitation, \\
& & & energy deposition, and production\\
& & &  of secondary electrons \\
Prechemical and chemical & $10^{-12}$--$10^{-6}$ s & $R$, $\Omega_R$, $G_R$, $G_{\rm NP}$ & Water radiolysis, radical diffusion, \\
& & & oxygen reactions, scavenging, quenching, \\
& & & and radical--radical recombination \\
Initial DNA damage & $10^{-12}$--$10^{-3}$ s & $N_0$, $S_0$, $C_0$, $f_c$ & Induction of base damage,  \\
& & & cSSBs, and simple, omplex,  \\
& & & or clustered DSBs\\
Damage recognition and signalling & seconds--hours & $A$, $H$, $P$ &
ATM activation, $\gamma$H2AX formation, \\
& & & chromatin signalling,  and recruitment of  \\ 
& & & end-recognition and repair factors\\
Pathway allocation & minutes--hours & $p_N$, $p_H$, $p_A$, $p_{\rm SSA}$, & Competition among c-NHEJ, HR, \\
& & $\mu_{js}(L,C_n)$  &alternative end joining,  MMEJ, and SSA; \\
& & &  cell-cycle-dependent pathway choice\\
Fast repair & minutes--hours & $S(t)$, $C_r(t)$, $\lambda_1$ & Rapid processing of simple lesions and the \\
& & & repairable complex-lesion component \\
Persistent repair & hours--days & $C_p(t)$, $\rho_C$, $\lambda_2$ & Slow processing or persistence of highly \\
& & & complex and clustered lesions \\ 
Binary end interaction & minutes--days & $\eta$, $\theta_0$, $\sigma$, $R_m(t)$ & Non-cognate end interaction, \\
& & & erroneous end joining, deletions, exchanges,  \\
& & & and chromosome-aberration precursors\\
Endpoint evaluation & $T=24$ h & $R_c(T)$, $R_m(T)$, $U(T)$ & Exclusive classification into\\
& & &  correct repair,  misrepair and  \\
& & & residual damage: $N_0=R_c+R_m+U$   \\  \hline
\end{tabular}
\end{table*}

The proton field is described by the differential fluence rate $\Phi_{p} (E, \bf{r},t)$, which specifies the number of protons with kinetic energy (E) traversing a unit area per unit time 
and energy interval at position $\mathbf r$. Together with the collision-stopping power $S_{\mathrm{col}}(E)$, this distribution determines the local absorbed-dose rate and the dose-averaged LET. 

To find the absorbed dose rate, we consider a small volume element $dV=dAdx$ with mass $dm=\rho dAdx$. The number of protons in the energy interval $[E,E+dE]$ crossing $dA$ during the time $dt$ is
\begin{displaymath}
dN_{p}= \Phi_{p}(E, {\bf{r}},t)dE dA dt.
\end{displaymath}
The energy deposited by each proton over the path length $dx$ is
\begin{displaymath}
d\epsilon_{p}= S_{col}(E)dx,
\end{displaymath}
where $S_{col} = -(dE/dx)_{col}$  gives the mean kinetic energy lost by a proton per unit length through ionisation and excitation of the medium. Therefore, the total energy deposited by these protons is
\begin{equation}
d\epsilon = \Phi_{p}(E, {\bf{r}}, t) S_{col}(E)dE dA dx dt
\end{equation}
Integrating over proton energy yields
\begin{displaymath}
d\epsilon = dA dx dt \int_{0}^{\infty} \Phi_{p}(E, {\bf{r}}, t) S_{col}(E) dE
\end{displaymath}
Since $dm= \rho dA dx$, 
\begin{displaymath}
\frac{dD_{p}}{dt} = \frac{1}{dm} \frac{d\epsilon}{dt},
\end{displaymath}
and consequently the absorbed-dose rate in water-equivalent material is
\begin{equation}
 \dot D_p(\mathbf r,t)=\frac{1}{\rho}\int \Phi_p(E,{\bf r},t)S_{\mathrm{col}}(E)d E,
 \label{eq:dose}
\end{equation}
Using the mass-stopping power, $\mathcal{S}_{\mathrm{m,col}} =  S_{\mathfrak{col)}}(E)/\rho$, the absorbed dose-energy rate becomes
\begin{equation}
 \dot D_p(\mathbf r,t)=\int_0^\infty  \Phi_p(E,{\bf r},t)\mathcal{S}_{\mathrm{m,col}}(E) dE.
 \label{eq:dose2}
\end{equation}
The accumulated dose at position ${\bf r}$ is
\begin{displaymath}
D_{p}({\bf r},t) = \int_0^t \dot D_{p}({\bf r},t')dt'.
\end{displaymath}
Therefore,
\begin{equation}
D_{p}({\bf r},t) = \frac{1}{\rho}\int_0^t dt' \int_0^\infty dE\Phi_{p}(E,{\bf r},t') S_{col}(E).
\end{equation}
If the time-integrated differential fluence is
\begin{equation}
F_{p}({\bf r},t) = \int_0^t \Phi_{p}(E,{\bf r},t'),
\end{equation}
then
\begin{equation}
D_{p}({\bf r},t) = \frac{1}{\rho} \int_0^\infty F_{p}(E,{\bf r},t') S_{col}(E) dE.
\end{equation} 
For proton energies relevant to radiotherapy, the collision-stopping power is generally the dominant contributor to local energy deposition. For greater transport accuracy, the separate inclusion of the nuclear interaction is required.
 The dose contribution from protons of energy $E$ is
\begin{displaymath}
d\dot{D}_{p}(E) = \frac{1}{\rho}\Phi_{p}(E)S_{col}(E)dE.
\end{displaymath}
The dose-averaged LET is defined by
\begin{displaymath}
L_{D}=\frac{\int L(E)d\dot{D}_{p}(E)}{\int d\dot{D}_{p}(E)}.
\end{displaymath}
Substituting the differential dose contribution, we obtain the dose-averaged LET as:
\begin{equation}
 L_D(\mathbf r,t)=
 \frac{\int \Phi_p S_{\mathrm{col}}^2 dE}
      {\int \Phi_p S_{\mathrm{col}} d E}.
 \label{eq:let}
\end{equation}
In terms of  the differential fluence rate, $\dot{F}_{p}$, the dose-averaged LET can be written as
\begin{equation}
L_{D}({\bf r},t)= \frac{\int_0^\infty \dot{F}_{p}(E,{\bf r},t)S^{2}_{col}(E)dE}{\int_0^\infty \dot{F}_{p}(E,{\bf r},t)S_{col}(E)dE},
\end{equation}

For $N_f$ fractions or pulses,
\begin{equation}
 \dot D_p(t)=\sum_{j=1}^{N_f}d_jg_j(t-t_j),\hspace{0.3cm}
 \int g_j(t) d t=1,
 \hspace{0.3cm} \sum_jd_j=D_{\mathrm{tot}}.
 \label{eq:pulses}
\end{equation}
The impulsive limit is obtained with $g_j(t)=\delta(t)$.

\subsection{Nanoparticle Dose and Distribution}

The nanoparticle distribution is represented by the spatially and temporally varying concentration $C_{\mathrm{NP}}(\mathbf r,t)$, with separate extracellular, 
cytoplasmic and nuclear or perinuclear compartments to account for cellular uptake, transport, clearance and localisation. Because nanoparticle-mediated enhancement is 
highly localised, the biological effect depends not only on the total intracellular nanoparticle concentration but also on nanoparticle size, composition, surface coating, 
aggregation state and distance from chromatin. A radial energy-deposition kernel $K_{\mathrm{NP}}(|\mathbf r-\mathbf r_i|;E_p,a_{\mathrm{NP}},Z_{\mathrm{NP}})$ is therefore 
assigned to each nanoparticle centred at $\mathbf r_i$, representing the local modification of proton slowing, secondary-electron production and radical generation. 
The resulting mixed-field dose is spatially heterogeneous and may be expressed as
\begin{displaymath}
D_{\mathrm{loc}}(\mathbf r)=D_p(\mathbf r)\left[1+\sum_i\eta_i  K_{\mathrm{NP}}(|\mathbf r-\mathbf r_i|;E_p,a_{\mathrm{NP}},Z_{\mathrm{NP}})\right].    
\end{displaymath}
where $a_{\mathrm{NP}}$  is the nanoparticle radius and $Z_{NP}$ is the atomic number of the nanoparticle material.  Consequently, two cells receiving the same macroscopic proton dose may experience different DNA damage per nucleus if their nanoparticle uptake or nuclear localisation differs. This spatial description provides the essential link between the incident proton spectrum, nanoparticle biodistribution, local LET enhancement and the subsequent induction of simple and clustered DNA lesions.

We consider a three-compartment minimal intracellular uptake model that separates extracellular $C_e$, cytoplasmic $C_c$, and nuclear $C_n$ concentrations \cite{Aberg2021,Hinderliter2010,Chithrani2006}.
Because the model assumes a perinuclear distribution of 15-nm AuNPs, it is more precise to replace $C_n$ in the transport equations by $C_{pn}$. 
 A volume-consistent first-order uptake and redistribution model is \cite{Aberg2021,Oh2011}
\begin{subequations}
\label{eq:perinuclear-uptake}
\begin{align}
\frac{dC_e}{dt}
={}&
I_{\mathrm{NP}}(t)
-\frac{V_c}{V_e}k_{\mathrm{in}}C_e
+\frac{V_c}{V_e}k_{\mathrm{out}}C_c
-k_{\mathrm{cl}}C_e,
\label{eq:Ce-pn}\\
\frac{dC_c}{dt}
={}&
k_{\mathrm{in}}C_e
-\left(k_{\mathrm{out}}+k_{c\mathrm{pn}}\right)C_c
+\frac{V_{\mathrm{pn}}}{V_c}
 k_{\mathrm{pn}c}C_{\mathrm{pn}},
\label{eq:Cc-pn}\\
\frac{dC_{\mathrm{pn}}}{dt}
={}&
\frac{V_c}{V_{\mathrm{pn}}}
 k_{c\mathrm{pn}}C_c
-\left(
k_{\mathrm{pn}c}+k_{\mathrm{deg,pn}}
\right)C_{\mathrm{pn}}.
\label{eq:Cpn}
\end{align}
\end{subequations}
where $V_{e}, V_{c}$ and $V_{pn}$ describe effective extracellular volume, cytoplasmic volume, and  perinuclear-shell volume, $k_{\mathrm{in}}$ and $k_{\mathrm{out}}$ describe cellular
uptake and exocytosis, respectively. The coefficients
$k_{c\mathrm{pn}}$ and $k_{\mathrm{pn}c}$ describe forward and reverse
transport between the cytoplasmic and perinuclear compartments,
$k_{\mathrm{cl}}$ is the extracellular clearance coefficient, and
$k_{\mathrm{deg,pn}}$ represents irreversible degradation,
sequestration, or loss from the perinuclear compartment. The volume
ratios ensure conservation of nanoparticle number when the state
variables are expressed as concentrations.

If the perinuclear compartment is represented by a shell surrounding an ellipsoidal nucleus with semiaxes $a_{n},b_{n},c_{n}$, then its effective volume may be approximated as
\begin{displaymath}
V_{pn} = \frac{4\pi}{3}\left[(a_{n}+\delta_{pn})(b_{n}+\delta_{pn})(c_{n}+\delta_{pn})-a_{n}b_{n}c_{n})\right]
\end{displaymath}
where $\delta_{pn}$  is the thickness of the perinuclear shell.

If the extracellular medium is sufficiently large that uptake does not appreciably change $C_{e}$, then $C_{e}(t)=C_{e,0}$ 
is an imposed boundary condition. In that case, $dC_{e}/dt=0$.  The cytoplasmic steady-state concentration is
\begin{equation}
C_{c}^{*} = \frac{k_{in}(k_{pnc}+k_{\mathrm{deg},pn})C_{e,0}}{(k_{pnc}k_{out}+k_{\mathrm{deg},pn}k_{out}+k_{\mathrm{deg,pn}k_{cpn}})}.
\end{equation}
The associated perinuclear steady-state concentration is
\begin{equation}
C_{pn}^{*} = \frac{k_{cpn}(k_{in}C_{e,0}}{(k_{pnc}k_{out}+k_{\mathrm{deg},pn}k_{out}+k_{\mathrm{deg,pn}k_{cpn}})}.
\end{equation}
If $C_{e}$ is not externally fixed, its kinetic equation must also be included. At steady state, the extracellular steady-state concentration is
\begin{equation}
C_{e}^{*} = \frac{I_{NP}\Delta}{(k_{in}+k_{cll})\Delta - k_{out}k_{in}(k_{pnc}+k_{\mathrm{deg},pn})},
\end{equation}
where
\begin{displaymath}
\Delta = (k_{pnc}+k_{\mathrm{deg},pn})(k_{out}+k_{cpn})-k_{pnc}k_{cpn}.
\end{displaymath}

When nanoparticles are introduced, the proton field generates additional local energy deposition through proton slowing, secondary electrons and nanoparticle-modified radical chemistry.
To calculate the excess dose associated with one nanoparticle, consider a nanoparticle centred at ${\bf r}_{i}$. Assume that its excess dose contribution is radially symmetric around its centre. 
The distance from the nanoparticle to the observation point ${\bf r}$  is $s_{i} = \mid {\bf r}-{\bf r}_{i}\mid$. 

The spatial distribution of GNPs  in the perinuclear region is described by
\begin{equation}
C_{pn} ({\bf r}) = \sum_{j=1}^{N_{NP} }\eta_{j} K_{NP}(s_{j}),
\label{eq:nanodistb}
\end{equation}
where $K_{NP}(s_{i})$ is a spatial localization kernel, and $\eta_{j}$ amounts for nanoparticle weighting, uptake or local accessibility.  
The kernel is normalised by $\int_VK_{NP}(r) d^3r=1$. A tractable screened form is
\begin{equation}
 K_{NP}(r)=A_K\frac{\exp(-r/l_e)}{r+r_c},
 \label{eq:kernel}
\end{equation}
where $A_K$ enforces normalisation, $\ell_e$ is an effective electron/radical range and $r_c$ regularises the origin. 

The excess dose generated by nanoparticle $i$ is
\begin{equation}
\Delta D_{i}({\bf r}) = D_{p}({\bf r})\eta_{i}K_{NP}(s_{i}),
\label{EqdoseNP}
\end{equation}
The relative dose enhancement produced by nanoparticle i is therefore
\begin{displaymath}
\frac{\Delta D_{i}({\bf r})}{D_{p}({\bf r})}= \eta_{i}K_{NP}(s_{i})
\end{displaymath}
For $N_{NP}$ nnoparticles centered at ${\bf r_{1}}$, ${\bf r_{2}}$, $\cdots$, the total excess dose, under linear-superposition approximation, is
\begin{displaymath}
\Delta D_{NP}({\bf r})= \sum_{i=1}^{N_{NP}}\Delta D_{i}({\bf r}).
\end{displaymath}
Substituting Eq. (\ref{EqdoseNP}) gives
\begin{equation}
\Delta D_{NP}({\bf r})= D_{p}({\bf r})\sum_{i=1}^{N_{NP}}\eta_{i}K_{NP}(s_{i}).
\label{EqdoseNP2}
\end{equation}
The total local dose is the sum of the background and excess doses:
\begin{displaymath}
D_{\mathrm{loc}}({\bf r}) = D_{p}({\bf r}) +\Delta D_{NP}({\bf r}),
\end{displaymath}
hence the local mixed-field dose is
\begin{equation}
 D_{\mathrm{loc}}(\mathbf r)=D_p(\mathbf r)  \left[1+\sum_{i=1}^{N_{NP}}\eta_i
 K_{NP}(|\mathbf r-{\mathbf r}_{i})\right].
 \label{eq:localdose}
\end{equation}
The quantity inside the square brackets is the local dose-enhancement factor, $\mathrm{DEF}$,
\begin{displaymath}
\mathrm{DEF}_{\mathrm{loc}}= \frac{D_{\mathrm{loc}}}{D_{p}({\bf r})}= \left[1+\sum_{i=1}^{N_{NP}}\eta_i
 K_{NP}(|\mathbf r-\mathbf r_i|;E_p,a,Z)\right].
\end{displaymath}

To first order,
\begin{equation}
\eta_j(L) \simeq \eta_{j0}+\eta_{jL},
\label{eq:eta}
\end{equation}
the local mixed-field dose becomes
\begin{eqnarray}
\langle D_{\mathrm{loc}}\rangle_{\mathrm{DNA}} & = & D_{p}\left[ 1+\Bigl\langle \sum_{j}\eta_{j0}K_{NP}\Bigl\rangle_{\mathrm{DNA}}  \right. \nonumber\\
& & \left. + L \Bigl\langle \sum_{j}\eta_{j0}K_{NP}\Bigl\rangle_{\mathrm{DNA}} \right].
\end{eqnarray}

\subsection{Reaction pathways of DNA-reactive chemical species}
\label{sec:reactive-species}

Proton irradiation generates short-lived electrons, radicals, and
molecular products through the radiolysis of the intracellular aqueous
medium. In the presence of AuNPs, additional reactive species may be
generated by secondary electrons and by the subsequent excitation and
ionization of nearby biomolecules. Because the AuNPs are assumed to
accumulate predominantly within a perinuclear region, the resulting
chemical source is spatially heterogeneous and depends on the local
perinuclear AuNP concentration.

\begin{figure*}[h]
\centering
\resizebox{0.98\textwidth}{!}{%
\begin{tikzpicture}[
  font=\small,
  processbox/.style={
    rectangle,
    rounded corners=3pt,
    draw=black,
    line width=0.9pt,
    align=center,
    text width=44mm,
    minimum height=20mm,
    inner xsep=4mm,
    inner ysep=3mm,
    fill=white
  },
  reaction/.style={
    -{Latex[length=3mm,width=2mm]},
    line width=1.1pt,
    shorten >=3pt,
    shorten <=3pt
  },
  numberlabel/.style={
    circle,
    draw=black,
    fill=white,
    line width=0.8pt,
    minimum size=6.5mm,
    inner sep=0pt,
    font=\bfseries\scriptsize
  }
]


\node[
  circle,
  draw=protonblue,
  fill=protonblue!10,
  line width=1.2pt,
  minimum size=16mm,
  inner sep=1mm
] (proton) at (0,0)
{\(\mathrm p^+\)};

\node[
  processbox,
  draw=protonblue,
  fill=waterblue,
  text width=40mm
] (water) at (4.6,0)
{
  \textbf{Water-equivalent}\\
  \textbf{cellular medium}\\[1.5mm]
  \(\mathrm{H_2O}\)
};

\node[
  processbox,
  draw=radicalred,
  fill=radicalred!7,
  text width=54mm,
  minimum height=22mm
] (radicals) at (11.3,0)
{
  \textbf{DNA-reactive species}\\[2mm]
  \(\mathrm{OH}^{\bullet},\;
    \mathrm H^{\bullet},\;
    e_{\mathrm{aq}}^{-},\;
    \mathrm O_2^{\bullet-},\;
    \mathrm{HO_2}^{\bullet}\)
};

\node[
  processbox,
  draw=dnagreen,
  fill=dnagreen!8,
  text width=50mm,
  minimum height=22mm
] (dna) at (18.3,0)
{
  \textbf{Initial DNA lesions}\\[2mm]
  base damage and SSBs\\
 simple, complex, and  clustered  DSBs\\
   $\displaystyle
    N_0=D_p \!\left[  Y_p(L)+ Y_{\mathrm{NP}}C_{pn}+ Y_{\mathrm{syn}}C_{pn}L \right]$
};


\node[
  circle,
  draw=goldNP!75!black,
  fill=goldNP!70,
  line width=1.2pt,
  minimum size=24mm,
  align=center,
  inner sep=1mm
] (np) at (5.1,-3.7)
{
  \(\mathrm{AuNP}\)\\[1mm]
  \(\scriptstyle d_{pn}=15 nm\)\\[1mm]
\( C_{pn}\)
};

\node[
  processbox,
  draw=black,
  fill=gray!8,
  text width=54mm,
  minimum height=24mm
] (loss) at (11.3,-3.8)
{
  \textbf{Radical-loss pathways}\\[2mm]
  scavenging and diffusion: \(k_qR\)\\[1mm]
  radical recombination: \(k_{RR}R^2\)
};


\draw[
  reaction,
  draw=protonblue
]
  (proton.east)
  --
  (water.west);

\node[numberlabel]
  at ($(proton.east)!0.5!(water.west)$)
  {1};

\draw[
  reaction,
  draw=protonblue
]
  (water.east)
  --
  (radicals.west);

\node[numberlabel]
  at ($(water.east)!0.5!(radicals.west)$)
  {2};

\draw[
  reaction,
  dashed,
  draw=goldNP!75!black
]
  (np.north east)
  to[
    out=35,
    in=235
  ]
  (radicals.south west);

\node[numberlabel]
  at (8.2,-2.25)
  {3};

\draw[
  reaction,
  draw=radicalred
]
  (radicals.east)
  --
  (dna.west);

\node[numberlabel]
  at ($(radicals.east)!0.5!(dna.west)$)
  {4};

\draw[
  reaction,
  draw=gray!80!black
]
  (radicals.south)
  --
  (loss.north);

\node[numberlabel]
  at ($(radicals.south)!0.5!(loss.north)$)
  {5};

\node[
  rectangle,
  rounded corners=3pt,
  draw=black!65,
  line width=0.8pt,
  align=left,
  text width=185mm,
  inner xsep=6mm,
  inner ysep=4mm,
  fill=blue!2
] (key) at (10.0,-7.5)
{
  \textbf{Reaction key}\\[2mm]
  \textbf{1.}\;
  Differential proton fluence:
  \(\Phi_p(E,\mathbf r,t)\).
  \qquad
  \textbf{2.}\;
Water radiolysis and production of DNA-reactive species: $G_R(L)\dot D_p(t)$. \\
  [2mm]
  \textbf{3.}\;
AuNP-mediated secondary-electron and radical production:
  \(
    G_{\mathrm{NP}}(L,C_{pn})\dot D_p(t),
    \qquad
    G_{\mathrm{NP}}(L,C_{pn})
    \simeq
    C_{pn}\!\left(G_0+G_LL\right).
  \)
  \\[2mm]
  \textbf{4.}\;
   Radical attack on DNA:
  $\displaystyle
    \frac{dN_{\mathrm{chem}}}{dt} =k_{RD}R$. The resulting damage is classified into base lesions,
  SSBs, and simple, complex, or clustered DSBs. \\
[2mm]
  \textbf{5.}\;
  Radical removal:
$\displaystyle
  k_qR+k_{RR}R^2$.
};
\end{tikzpicture}%
}

\caption{%
Production and competing reaction pathways of DNA-reactive chemical
species following proton irradiation in the presence of 15-nm
perinuclear GNPs. 
}
\label{Fig1}
\end{figure*}
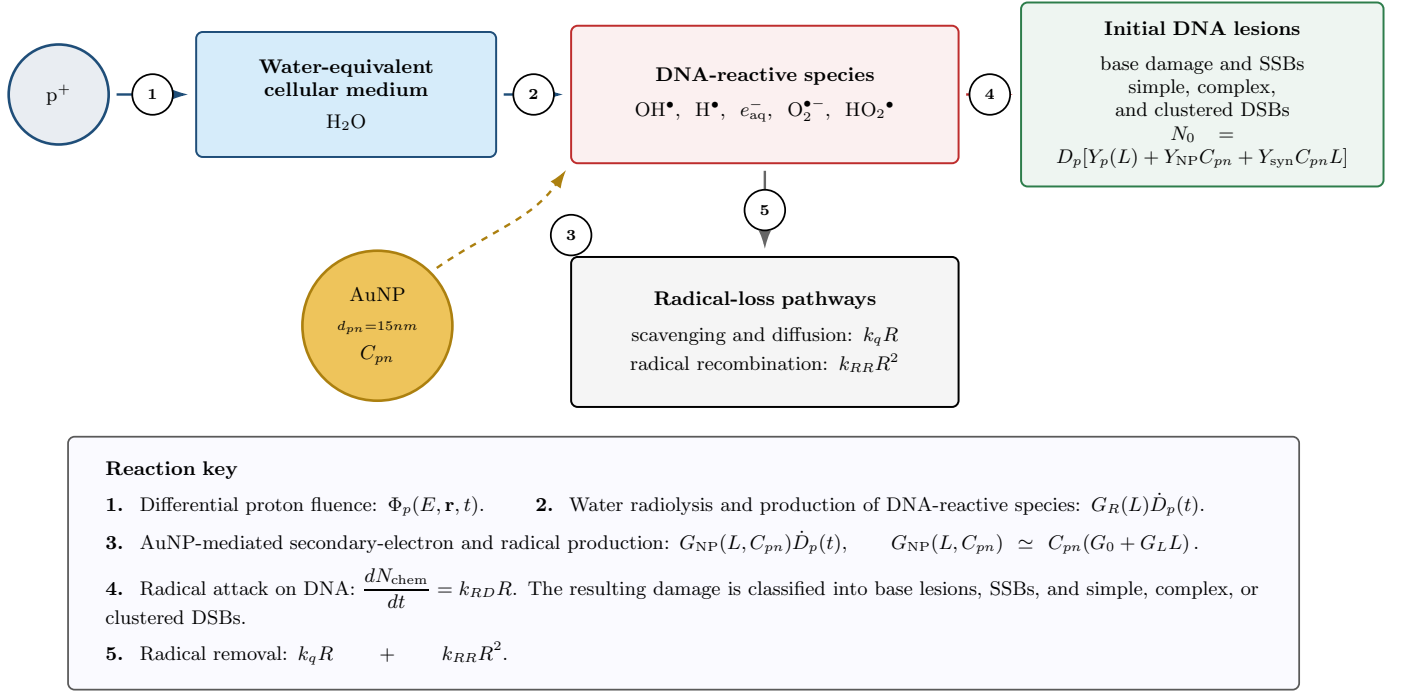

Let $R(\mathbf r,t)$ denote an effective concentration of DNA-reactive
chemical species. The $R$ may represent a weighted combination of species such as
$\mathrm{OH}^{\bullet}, \mathrm{H}^{\bullet}, \mathrm{e_{aq}^{-}}, \mathrm{O_{2}}^{\bullet -}, \mathrm{HO_{2}}^{\bullet}$ and longer-lived nanoparticle-generated oxidents (Fig. \ref{Fig1}). 
Proton transport and water radiolysis generate reactive species that either attack DNA or undergo scavenging, diffusion, and radical--radical recombination. The nanoparticle-mediated source depends on both the relative GNP concentration $C_n$ and proton LET $L$, producing the interaction represented macroscopically by the $C_nL$ terms in the initial-lesion yield and survival models. Initial DNA damage is classified into base damage, single-strand breaks, and simple, complex, or clustered double-strand breaks before repair-pathway allocation.

The spatial and temporal evolution of the radiation-generated
DNA-reactive species is described using a reduced reaction--diffusion
balance,
\begin{equation}
\frac{\partial R(\mathbf r,t)}{\partial t} = D_R\nabla^2R(\mathbf r,t)+S_R(\mathbf r,t) -k_qR(\mathbf r,t) -k_{RR}R^2(\mathbf r,t),
\label{eq:radical-rd}
\end{equation}
where $D_R$ is an effective diffusion coefficient, $k_q$ represents
first-order quenching, scavenging, and chemical decay, and $k_{RR}$
is the second-order radical--radical recombination coefficient.  This formulation follows established diffusion--reaction descriptions
of the chemical stage of water radiolysis and radiation-induced indirect DNA damage \cite{Plante2015,Peukert2019,Tran2021}.
The present model extends this framework by introducing a spatially dependent AuNP-mediated source term, motivated by nanoscale simulations
of AuNP-enhanced radiolytic-species production and DNA damage \cite{Klapproth2021}.

The local production rate is written as the product of the local proton dose rate and the reactive-species yield per unit dose:
\begin{displaymath}
S_{R}({\bf r},t) = G_{tot}({\bf r},L,po_{2},t)\dot{D}_{p}({\bf r},t) 
\end{displaymath}
The total yield is decomposed into three contributions:
\begin{displaymath}
G_{tot}=G_{p}(L,po_{2})+G_{NP}G_{pn}({\bf r},t)+G_{syn}C_{pn}({\bf r},t)L,
\end{displaymath}
where
$G_{p}(L,po_{2})$ is the proton-induced yield without AuNPs, $G_{NP}G_{pn}({\bf r},t)$ represents the LET-independent AuNP enhancement, and $G_{syn}C_{pn}({\bf r},t)L$ describes the  coupled AuNP–LET contribution. Consequently,
\begin{eqnarray}
S_{R}({\bf r},t) &= & \left[ G_{p}(L,po_{2})+G_{NP}G_{pn}({\bf r},t)\right.\nonumber\\
& & \left. +G_{syn}C_{pn}({\bf r},t)L  \right] \dot{D}_{p}({\bf r},t) 
\label{eq:localprod}
\end{eqnarray}

The oxygen dependence of the proton-induced component is represented by\cite{Wenzl2011,Grimes2015}
\begin{equation}
G_p(L,p_{\mathrm O_2})
=
G_{\mathrm{ind}}(L)
+
G_{\mathrm{ox}}(L)
\frac{p_{\mathrm O_2}}
{K_{\mathrm O_2}+p_{\mathrm O_2}},
\label{eq:oxygen-radical-yield}
\end{equation}
where $G_{\mathrm{ind}}$ is the oxygen-independent contribution,
$G_{\mathrm{ox}}$ is the maximum oxygen-mediated contribution, and
$K_{\mathrm O_2}$ is the oxygen partial pressure producing one-half
of the maximum oxygen-dependent yield.

A DNA-weighted spatial average of Eq.~(\ref{eq:radical-rd}), together with the well-mixed closure $\overline{R^2}\simeq\overline R^{\,2}$, gives the reduced kinetic
equation
\begin{eqnarray}
\frac{dR}{dt} &= & \left[ G_p(L,p_{\mathrm O_2}) +G_{\mathrm{NP}}C_{\mathrm{pn}}^{\mathrm{eff}} +G_{\mathrm{syn}}C_{\mathrm{pn}}^{\mathrm{eff}}L \right]\dot D_p(t) \nonumber\\
& &  -k_qR-k_{RR}R^2,
\label{eq:radical-riccati}
\end{eqnarray}
where $C_{\mathrm{pn}}^{\mathrm{eff}}$ is the DNA-weighted effective
perinuclear AuNP concentration. The quadratic loss term produces
chemical saturation: increasing the prompt production rate does not
produce a proportional increase in the time-integrated reactive-species
exposure.

For a sufficiently short irradiation pulse, radical loss during the
pulse may be neglected. The concentration immediately after the pulse
is then
\begin{equation}
R_0
=
D_p
\left[
G_p(L,p_{\mathrm O_2})
+G_{\mathrm{NP}}C_{\mathrm{pn}}^{\mathrm{eff}}
+G_{\mathrm{syn}}C_{\mathrm{pn}}^{\mathrm{eff}}L
\right].
\label{eq:postpulse-R0}
\end{equation}
After termination of the pulse, the source vanishes and
Eq.~(\ref{eq:radical-riccati}) reduces to
$dR/dt=-k_qR-k_{RR}R^2$. With $R(0^+)=R_0$, its solution is
\begin{equation}
R(t)
=
\frac{R_0e^{-k_qt}}
{1+\left(k_{RR}R_0/k_q\right)
\left(1-e^{-k_qt}\right)}.
\label{eq:postpulse-R}
\end{equation}
The corresponding time-integrated reactive-species exposure is
\begin{equation}
\Omega_R
=
\int_0^\infty R(t)\,dt
=
\frac{1}{k_{RR}}
\ln\left(
1+\frac{k_{RR}R_0}{k_q}
\right).
\label{eq:radical-exposure}
\end{equation}

\subsection{Track-to-track Heterogeneity and Mixed-field Yield}
Lesion intensity is modelled as a Gamma-distributed latent variable to account for unresolved heterogeneity across individual nuclear traversals \cite{Loan2022}. While lesion formation, conditional on a specified track geometry, can be approximated by a Poisson process, the mean intensity varies across traversals due to factors such as track length, proton energy, LET, chromatin density, oxygenation, radical scavenging, and nanoparticle proximity. The Gamma distribution offers a positive and flexible two-parameter framework to represent this variability, with mean $\alpha /\beta$, variance $\alpha /\beta^{2}$, and squared coefficient of variation $1/\alpha$. Marginalising the conditional Poisson count over the Gamma-distributed intensity yields a negative-binomial lesion distribution that accounts for overdispersion and excess zero counts not captured by a homogeneous Poisson model. The Gamma assumption serves as an effective population-level approximation rather than a fundamental law of proton-track physics. Its suitability should be evaluated using track-level lesion-count data and alternative positive mixing distributions. The lesion intensity $\Lambda$ need not follow a Gamma distribution as a fundamental law; instead, the Gamma distribution provides a physically motivated statistical model for unresolved variation in the mean lesion yield across nuclear traversals. The central assumption is
\begin{displaymath}
K\mid \Lambda = \lambda \approx \mathrm{Poisson}(\lambda )
\end{displaymath}
where $K$ is the number of lesions produced by a traversal, and $\Lambda$ is that traversal’s latent mean lesion intensity. The Gamma distribution is then assigned to $\Lambda$ to represent heterogeneity 
in track geometry, LET, chromatin density, nanoparticle proximity and radical chemistry.

The Gamma density in the shape-rate convention is:
\begin{equation}
 f(\Lambda)=\frac{\beta^\alpha}{\Gamma(\alpha)}
 \Lambda^{\alpha-1}e^{-\beta\Lambda}.
\end{equation}
Conditional on $\Lambda$, the DSB count is Poisson. Marginalisation gives 
\begin{align}
 P(K=k)
 &=\int_0^\infty P(K=k\mid\Lambda)f(\Lambda) d\Lambda\nonumber\\
 &=\frac{\Gamma(k+\alpha)}{\Gamma(\alpha)k!}  \left(\frac{\beta}{\beta+1}\right)^\alpha  \left(\frac{1}{\beta+1}\right)^k.
 \label{eq:nb}
\end{align}
The mean and variance follow from total expectation and total variance:
\begin{equation}
 \mathbb E[K]=\frac{\alpha}{\beta},\qquad
 \operatorname{Var}(K)=\frac{\alpha}{\beta}+\frac{\alpha}{\beta^2}.
 \label{eq:nbmoments}
\end{equation}
The zero-lesion and productive-track probabilities are
\begin{equation}
 P_0=\left(\frac{\beta}{\beta+1}\right)^\alpha,
 \qquad q=1-P_0.
 \label{eq:pzero}
\end{equation}
If $n$ proton traversals intersect a nucleus, the expected number of DSB-producing tracks is $n_p=nq$. For $N_0$ initial DSBs, the conditional mean multiplicity among productive tracks is
\begin{equation}
 \lambda_p=\frac{N_0}{n_p}.
 \label{eq:lambdap}
\end{equation}
Therefore, the macroscopic yield equation and microscopic Poisson–Gamma model cannot be calibrated independently: their means must agree.
The expected initial DSB number is decomposed into proton, nanoparticle and interaction terms (see appendix \ref{app:Yield} for details):
\begin{equation}
 N_0=D_p\left[Y_p(L)+Y_{NP}C_n+Y_{\mathrm{syn}}C_nL\right],
 \label{eq:yield}
\end{equation}
 where $Y_{p}(L)$ is the proton-only DSB yield, $Y_{NP}C_{n}$ describes the radiation-mediated nanoparticle contribution, and $Y_{\mathrm{syn}}C_{n}L$ represents the 
LET-dependent proton–nanoparticle interaction. This expression follows from a first-order expansion of the mixed-field yield in nanoparticle concentration, with 
the concentration sensitivity subsequently expanded to first order in LET. The interaction coefficient $Y_{\mathrm{mix}}$
 is therefore the mixed derivative 
\begin{displaymath}
\frac{\partial^{2}Y_{\mathrm{mix}}}{\partial C_{n}\partial L}
\end{displaymath}
 and quantifies the extent to which nanoparticle enhancement changes with proton LET. All contributions vanish at zero proton dose; consequently, nanoparticle-only toxicity, if present, 
must be introduced as a separate additive term. Equation (27) should be regarded as a low-concentration, moderate-dose response-surface approximation whose parameters 
require simultaneous calibration across multiple LET values and nanoparticle concentrations.

An equivalent factorised expression is
\begin{equation}
 N_0=D_pY_p(L)\left[1+\epsilon_{NP}C_n\mathcal F(L)\right].
\end{equation}
The cross term $Y_{\mathrm{syn}}$ is identifiable only when experiments include several LET values and intracellular nanoparticle concentrations.

\begin{figure*}[h]
\centering

\resizebox{0.80\textwidth}{!}{%
\begin{tikzpicture}[
    font=\small,
    >=Latex,
    node distance=7mm and 10mm,
    box/.style={
        rectangle,
        rounded corners=2pt,
        draw,
        line width=0.8pt,
        align=center,
        inner sep=6pt,
        minimum height=13mm
    },
    input/.style={
        box,
        draw=blue!75!black,
        fill=blue!6,
        minimum width=61mm
    },
    process/.style={
        box,
        draw=violet!75!black,
        fill=violet!6,
        minimum width=132mm
    },
    burden/.style={
        box,
        draw=blue!75!black,
        fill=blue!8,
        minimum width=132mm
    },
    simple/.style={
        box,
        draw=blue!80!black,
        fill=blue!7,
        minimum width=61mm
    },
    complex/.style={
        box,
        draw=orange!85!black,
        fill=orange!12,
        minimum width=61mm
    },
    selection/.style={
        box,
        draw=red!60!black,
        fill=red!5,
        minimum width=132mm,
        minimum height=25mm
    },
    correct/.style={
        box,
        draw=green!60!black,
        fill=green!8,
        minimum width=61mm
    },
    misrepair/.style={
        box,
        draw=red!80!black,
        fill=red!7,
        minimum width=61mm
    },
    residual/.style={
        box,
        draw=black!65,
        fill=black!4,
        minimum width=132mm
    },
    arrow/.style={
        -{Latex[length=2.4mm,width=1.7mm]},
        line width=0.9pt
    },
    bluearrow/.style={
        arrow,
        draw=blue!80!black
    },
    orangearrow/.style={
        arrow,
        draw=orange!90!black
    },
    greenarrow/.style={
        arrow,
        draw=green!55!black
    },
    redarrow/.style={
        arrow,
        draw=red!80!black
    },
    grayarrow/.style={
        arrow,
        draw=black!65
    },
    reaction/.style={
        circle,
        draw=black,
        fill=white,
        line width=0.7pt,
        minimum size=7mm,
        inner sep=0pt,
        font=\footnotesize
    }
]


\node[input] (proton) {
    \textbf{Proton irradiation}\\[1mm]
    dose $D_p$, dose rate $\dot D_p(t)$,\\
    LET $L$, oxygen pressure $p_{\mathrm O_2}$
};

\node[input, right=12mm of proton] (aunp) {
    \textbf{Perinuclear AuNP distribution}\\[1mm]
    $C_{\mathrm{pn}}(\mathbf r,t)$,
    \quad $C_{\mathrm{pn}}^{\mathrm{eff}}$
};


\node[process, below=10mm of $(proton.south)!0.5!(aunp.south)$]
(local) {
    \textbf{Local physical and chemical enhancement}\\[1.5mm]
    $\displaystyle
    D_{\mathrm{loc}}(\mathbf r)
    =
    D_p(\mathbf r)
    \left[
    1+
    \sum_{j=1}^{N_{\mathrm{NP}}}
    \eta_j(L)
    K_{\mathrm{NP}}\!\left(
    |\mathbf r-\mathbf r_j|
    \right)
    \right]
    $
    \\[2mm]
    $\displaystyle
    \frac{\partial R}{\partial t}
    =
    D_R\nabla^2R
    +
    S_R(\mathbf r,t)
    -k_qR-k_{RR}R^2
    $
    \qquad
    $\displaystyle
    \Omega_R=\int_0^\infty R(t)\,dt
    $
};

\draw[bluearrow]
(proton.south) -- node[reaction,pos=0.52] {1} (local.north west);

\draw[orangearrow]
(aunp.south) -- node[reaction,pos=0.52] {1} (local.north east);


\node[burden, below=8mm of local] (burden) {
    \textbf{Initial DSB burden}\\[1.5mm]
    $\displaystyle
    N_0 =  D_p
    \left[
    Y_p(L)
    +
    Y_{\mathrm{NP}}C_{\mathrm{pn}}^{\mathrm{eff}}
    +
    Y_{\mathrm{syn}}
    C_{\mathrm{pn}}^{\mathrm{eff}}L
    \right],
    \qquad
    N_0=S_0+C_0
    $
};

\draw[bluearrow]
(local.south) -- node[reaction,pos=0.50] {2} (burden.north);


\node[simple, below left=9mm and 5mm of burden.south] (simpleDSB) {
    \textbf{Simple DSBs}\\[1mm]
    $\displaystyle S_0=N_0(1-f_c)$\\
    isolated or readily ligatable ends
};

\node[complex, below right=9mm and 5mm of burden.south] (complexDSB) {
    \textbf{Complex DSBs}\\[1mm]
    $\displaystyle C_0=N_0f_c$\\
    clustered DSBs, damaged termini,\\
    or associated base lesions
};

\draw[bluearrow]
(burden.south) -- ++(0,-4mm)
-| node[reaction,pos=0.73] {3} (simpleDSB.north);

\draw[orangearrow]
(burden.south) -- ++(0,-4mm)
-| node[reaction,pos=0.73] {3} (complexDSB.north);

\node[
    align=center,
    font=\footnotesize,
    below=3.5mm of burden.south
] {
    $\displaystyle
    f_c=f_c\!(
    L,C_{\mathrm{pn}}^{\mathrm{eff}},\Omega_R)$
};


\node[selection, below=13mm of $(simpleDSB.south)!0.5!(complexDSB.south)$]
(selection) {
    \textbf{Repair-pathway selection}\\[1.5mm]
    \begin{tabular}{ccc}
    c-NHEJ &
    HR/resection &
    alternative end joining \\[-1mm]
    $p_{N,S},\,p_{N,C}$ &
    $p_{H,S},\,p_{H,C}$ &
    $p_{A,S},\,p_{A,C}$
    \end{tabular}
    \\[2mm]
    $\displaystyle
    p_{N,x}+p_{H,x}+p_{A,x}=1,
    \qquad x\in\{S,C\}
    $
};

\draw[bluearrow]
(simpleDSB.south) -- node[reaction,pos=0.52] {4} (selection.north west);

\draw[orangearrow]
(complexDSB.south) -- node[reaction,pos=0.52] {4} (selection.north east);


\node[correct, below left=10mm and 5mm of selection.south] (correctrepair) {
    \textbf{Correct repair}\\[1mm]
    correct-end joining or\\
    homologous restoration\\[1mm]
    $\displaystyle R_c(t)$
};

\node[misrepair, below right=10mm and 5mm of selection.south] (misrepair) {
    \textbf{Misrepair}\\[1mm]
    incorrect-end joining, deletion,\\
    translocation, or aberration\\[1mm]
    $\displaystyle R_m(t)$
};

\draw[greenarrow]
(selection.south) -- ++(0,-4mm)
-| node[reaction,pos=0.72] {5}
node[above left=3mm,pos=0.80,font=\footnotesize]
{$\mu_{j,x}$}
(correctrepair.north);

\draw[redarrow]
(selection.south) -- ++(0,-4mm)
-| node[reaction,pos=0.72] {6}
node[above right=2mm,pos=0.80,font=\footnotesize]
{$1-\mu_{j,x}$}
(misrepair.north);


\node[residual, below=13mm of $(correctrepair.south)!0.5!(misrepair.south)$]
(residual) {
    \textbf{Residual damage}\\[1mm]
    unrepaired or slowly repaired simple and complex DSBs\\[1mm]
    $\displaystyle U(t)=S(t)+C(t)$
};

\coordinate (leftlane) at ($(simpleDSB.west)+(-8mm,0)$);

\draw[grayarrow]
(simpleDSB.west)
-- (leftlane)
|- node[reaction,pos=0.88] {7}
(residual.west);

\coordinate (rightlane) at ($(complexDSB.east)+(8mm,0)$);

\draw[grayarrow]
(complexDSB.east)
-- (rightlane)
|- node[reaction,pos=0.88] {7}
(residual.east);

\end{tikzpicture}%
}

\vspace{2mm}
\fbox{
\begin{minipage}{0.92\textwidth}
\small
\textbf{Reaction key.}
\textcircled{\scriptsize 1}
Proton irradiation and the spatially heterogeneous perinuclear AuNP
distribution determine the local physical and chemical environment.\\
\textcircled{\scriptsize 2}
Local energy deposition and reactive-species production establish the
initial DSB burden.\\
\textcircled{\scriptsize 3}
The initial lesions are partitioned into simple and complex DSBs through
the clustered-lesion fraction $f_c$.\\
\textcircled{\scriptsize 4}
Simple and complex DSBs enter competing c-NHEJ, HR/resection, and
alternative end-joining pathways.\\
\textcircled{\scriptsize 5}
A pathway- and lesion-dependent fraction $\mu_{j,x}$ undergoes correct
repair.\\
\textcircled{\scriptsize 6}
The complementary fraction $1-\mu_{j,x}$ undergoes misrepair.\\
\textcircled{\scriptsize 7}
Lesions that remain unrepaired or are repaired slowly constitute the
residual damage $U(t)=S(t)+C(t)$.
\end{minipage}
}
\caption{
Multiscale pathway model for proton--AuNP-induced DNA damage
and repair. }
\label{fig:dna-repair-pathways}
\end{figure*}

\section{Damage recognition, signalling, and pathway allocation}
\label{sec:damage-signalling-allocation}

DNA-damage recognition and signalling are conceptualised as a sequential, saturable response that links the time-dependent populations of simple and complex double-strand breaks (DSBs) to ATM activation, $\gamma$H2AX formation, recruitment of DNA-end-recognition and processing factors, and allocation among competing repair pathways \cite{Tan2023,Vitti2019,Souli2022}. Figure \ref{fig:dna-repair-pathways} illustrates how proton dose, dose rate, LET, oxygenation, and the spatially heterogeneous perinuclear AuNP distribution collectively determine local physical-dose enhancement and the production of DNA-reactive chemical species. These processes establish the initial DSB lesions per nucleus $N_0$, which is subsequently partitioned into simple and complex lesions according to the clustered-lesion fraction $f_c$, with initial simple and complex DSB populations given by
\begin{equation}
S_0=N_0(1-f_c),
\qquad
C_0=N_0f_c,
\label{eq:SC0-signalling-allocation}
\end{equation}
where
\begin{equation}
f_c =
f_c\left(
L,C_{\mathrm{pn}}^{\mathrm{eff}},\Omega_R
\right).
\label{eq:fc-signalling-allocation}
\end{equation}

Simple double-strand breaks (DSBs) generally possess clean, readily ligatable termini and minimal collateral DNA damage. These lesions are primarily recognised and repaired via canonical nonhomologous end joining (c-NHEJ). In contrast, complex DSBs are characterised by clustered damage, including additional strand breaks, damaged bases, abasic sites, or chemically modified termini within a short DNA segment. Processing of complex DSBs typically requires end cleaning or resection prior to ligation, which may facilitate homologous recombination, resection-dependent repair, or alternative end joining. The need for more extensive processing and the prolonged persistence of complex lesions increase the likelihood of incomplete repair or non-cognate end interaction, thereby elevating the risk of deletions, translocations, chromosome aberrations, and cell death. The complexity fraction $f_c$ is determined by radiation quality, such as proton LET, and by nanoparticle-mediated changes in local ionisation and radical density. As a result, treatment-induced increases in $f_c$ can alter biological effectiveness even if the total initial DSB lesions per nucleus remain unchanged. Upstream physical and chemical variables are integrated into the signalling and repair stages through $N_0$, $f_c$, $S_0$, and $C_0$, and are not reintroduced as independent forcing terms. This approach prevents double counting of the AuNP-mediated enhancement.

Complex DSBs generally produce a stronger and more persistent damage-response signal than readily ligatable simple DSBs. The
effective lesion burden driving the signalling module is therefore defined as
\begin{equation}
B(t)=S(t)+\omega_CC(t),
\qquad
\omega_C\geq1,
\label{eq:effective-burden-allocation}
\end{equation}
where $\omega_C$ is a phenomenological complexity weight. For
$\omega_C=1$, the signalling burden reduces to the total residual DSB
population $U(t)=S(t)+C(t)$.

\subsection{ATM--$\gamma$H2AX signalling and factor recruitment}

Let $A(t)$ denote the activated fraction of the available ATM pool,
$H_\gamma(t)$ the normalized $\gamma$H2AX response, and $P(t)$ the
normalized recruitment of DSB-recognition and end-processing factors.
The symbol $H_\gamma$ is used to distinguish the $\gamma$H2AX signal
from the repair-pathway label $H$, which denotes HR/resection-mediated
repair  \cite{Bakkenist2003,Burma2001,Stucki2005,Meyer2013}.

ATM activation is driven by the Hill response
\begin{equation}
u_B(B)
=
\frac{B^{h_A}}
{K_A^{h_A}+B^{h_A}},
\label{eq:burden-Hill-allocation}
\end{equation}
where $K_A$ is the effective half-activation burden and $h_A$ controls
the steepness of ATM activation. The sequential signalling system is
\begin{subequations}
\label{eq:signalling-allocation-system}
\begin{align}
\frac{dA}{dt}
={}&
k_Au_B[B(t)](1-A)-k_{-A}A,
\label{eq:ATM-allocation}
\\
\frac{dH_\gamma}{dt}
={}&
k_HA(1-H_\gamma)-k_{-H}H_\gamma,
\label{eq:Hgamma-allocation}
\\
\frac{dP}{dt}
={}&
k_PH_\gamma(1-P)-k_{-P}P.
\label{eq:P-allocation}
\end{align}
\end{subequations}
The factors $1-A$, $1-H_\gamma$, and $1-P$ represent the finite
inactive or unrecruited pools. The coefficients $k_{-A}$, $k_{-H}$,
and $k_{-P}$ describe ATM deactivation, $\gamma$H2AX signal removal,
and dissociation or turnover of the recruited machinery, respectively.
The enhanced weight assigned to complex lesions is consistent with the stronger ATM- and DNA-PK-dependent $\gamma$H2AX response reported following clustered DNA 
damage \cite{Meyer2013}. The Hill activation term and finite-pool factors are modelling assumptions introduced here to represent threshold-like activation, saturation, 
and signal attenuation and should not be interpreted as equations taken directly from those studies.

For a lesion burden that varies slowly compared with the signalling
relaxation times, the instantaneous quasi-steady responses are
\begin{subequations}
\label{eq:quasisteady-signalling-allocation}
\begin{align}
A^*(B)
={}&
\frac{k_Au_B(B)}
{k_Au_B(B)+k_{-A}},
\\
H_\gamma^*(B)
={}&
\frac{k_HA^*(B)}
{k_HA^*(B)+k_{-H}},
\\
P^*(B)
={}&
\frac{k_PH_\gamma^*(B)}
{k_PH_\gamma^*(B)+k_{-P}}.
\end{align}
\end{subequations}

\subsection{Allocation among competing repair pathways}
\label{sec:pathway-allocation}

Following damage recognition and recruitment, each DSB is allocated
among three competing repair routes: canonical non-homologous end
joining (c-NHEJ), denoted by $N$; homologous recombination or
resection-mediated repair, denoted by $H$; and alternative end joining,
denoted by $A$. Because simple and complex DSBs differ in end
structure, chromatin context, and processing requirements, pathway
allocation is defined separately for each lesion class
$x\in\{S,C\}$.

To distinguish pathway selection from the rate of subsequent repair,
let $\kappa_{j,x}$ denote the intrinsic allocation propensity of
pathway $j$ for lesion class $x$. Motivated by the established competition among c-NHEJ,
HR/resection-mediated repair, and alternative end joining, and by the
dependence of pathway choice on lesion structure, factor recruitment,
end resection, and cell-cycle state
\cite{Chang2017,Ceccaldi2016,Scully2019}, we introduce
lesion-class-dependent, unnormalized pathway weights.

The corresponding unnormalized
pathway weights are defined as
\begin{subequations}
\label{eq:pathway-weights}
\begin{align}
w_{N,x}(t)
={}&
\kappa_{N,x}\,\Phi_N[P(t)],
\label{eq:N-weight}
\\
w_{H,x}(t)
={}&
\kappa_{H,x}\,
g_{\mathrm{HR}}(t)\,
\Phi_H[P(t)],
\label{eq:H-weight}
\\
w_{A,x}(t)
={}&
\kappa_{A,x}\,\Phi_A[P(t)].
\label{eq:A-weight}
\end{align}
\end{subequations}
Here, $\Phi_j(P)$ describes the dependence of pathway engagement on
the recruited processing machinery, and $g_{\mathrm{HR}}(t)$ is the
fractional HR-competence factor. The latter accounts for the
availability of a sister chromatid and the cell-cycle dependence of
resection-mediated repair.

A bounded recruitment function may be written as
\begin{equation}
\Phi_j(P)
=
\varepsilon_j
+
(1-\varepsilon_j)
\frac{P^{m_j}}
{K_{P,j}^{m_j}+P^{m_j}},
\qquad j\in\{N,H,A\},
\label{eq:pathway-recruitment-function}
\end{equation}
where $K_{P,j}$ is the half-activation recruitment level, $m_j$ is a
Hill coefficient, and $\varepsilon_j$ represents basal pathway
accessibility. Setting $\varepsilon_j=0$ gives a pathway that requires
recruitment, whereas $\varepsilon_j>0$ permits limited basal pathway
engagement.

At the population level, a minimal HR-competence factor is
\begin{equation}
g_{\mathrm{HR}}(t)=\phi_{S/G_2}(t),
\label{eq:HR-cell-cycle}
\end{equation}
where $\phi_{S/G_2}$ is the fraction of cells possessing an available
sister chromatid. If resection competence is modelled explicitly, one
may instead use
\begin{equation}
g_{\mathrm{HR}}(t)
=
\phi_{S/G_2}(t)
\frac{Q(t)^{m_H}}
{K_{\mathrm{HR}}^{m_H}+Q(t)^{m_H}},
\label{eq:HR-competence}
\end{equation}
where $Q(t)$ is a normalized measure of end-resection competence.

Following the standard competing-channel propensity construction
\cite{Gillespie1977}, the normalized allocation probabilities are obtained from the
competing pathway weights:

\begin{equation}
p_{j,x}(t)
=
\frac{w_{j,x}(t)}
{\mathcal W_x(t)},
\qquad
j\in\{N,H,A\},
\quad
x\in\{S,C\},
\label{eq:pathway-probability-general}
\end{equation}
where
\begin{equation}
\mathcal W_x(t)
=
w_{N,x}(t)+w_{H,x}(t)+w_{A,x}(t).
\label{eq:pathway-normalization-weight}
\end{equation}
Explicitly,
\begin{subequations}
\label{eq:pathway-probabilities}
\begin{align}
p_{N,x}(t)
={}&
\frac{
\kappa_{N,x}\Phi_N[P(t)]
}{
\mathcal W_x(t)
},
\\
p_{H,x}(t)
={}&
\frac{
\kappa_{H,x}g_{\mathrm{HR}}(t)\Phi_H[P(t)]
}{
\mathcal W_x(t)
},
\\
p_{A,x}(t)
={}&
\frac{
\kappa_{A,x}\Phi_A[P(t)]
}{
\mathcal W_x(t)
}.
\end{align}
\end{subequations}
By construction,
\begin{equation}
p_{N,x}(t)+p_{H,x}(t)+p_{A,x}(t)=1.
\label{eq:pathway-probability-normalization}
\end{equation}

For a minimal reduction corresponding closely to the original
allocation model, one may set
$\Phi_N(P)=P$ and $\Phi_H(P)=\Phi_A(P)=1$. The allocation
probabilities then become \cite{Fattah2010,Shibata2011,EscribanoDiaz2013}
\begin{subequations}
\label{eq:minimal-allocation}
\begin{align}
p_{N,x}
={}&
\frac{
\kappa_{N,x}P
}{
\kappa_{N,x}P
+\kappa_{H,x}g_{\mathrm{HR}}
+\kappa_{A,x}
},
\\
p_{H,x}
={}&
\frac{
\kappa_{H,x}g_{\mathrm{HR}}
}{
\kappa_{N,x}P
+\kappa_{H,x}g_{\mathrm{HR}}
+\kappa_{A,x}
},
\\
p_{A,x}
={}&
\frac{
\kappa_{A,x}
}{
\kappa_{N,x}P
+\kappa_{H,x}g_{\mathrm{HR}}
+\kappa_{A,x}
}.
\end{align}
\end{subequations}

For the late ${\mathrm{S/G2}}$, the single-strand annealing (SSA) contributes to the probabilities through
\begin{equation}
p_{j,x} = \frac{w_{j,x}}{w_{N,x}+w_{H,x}+w_{A,x}+w_{SSA,x}},  \qquad j\in \{ N,H,A,SSA \}
\end{equation}
The LET–GNP dependence is incorporated through an effective complexity variable 
\begin{displaymath}
\chi (L, C_{pn}^{\mathrm{eff}}) = \chi_{0}+\chi_{L}L+\chi_{C}C_{pn}^{\mathrm{eff}}+\chi_{LC}LC_{pn}^{\mathrm{eff}},
\end{displaymath}
which modifies the pathway weights. A transparent parametrization is
\begin{eqnarray}
w_{N} & = & w_{N}^{(0)}e^{-a_{N}\chi}, \qquad w_{H}=w_{H}^{(0)}g_{HR}e^{a_{H\chi }-b_{H}\chi^{2}}\nonumber\\
w_{A} & = &w_{A}^{(0)}e^{a_{A}\chi}, \qquad  \qquad  w_{SSA}  =  w_{SSA}^{(0)}g_{SSA}e^{a_{SSA}\chi },\nonumber
\end{eqnarray}
 It is expected that increasing LET and GNP concentration suppress c-NHEJ because increasingly complex ends are less readily bound and ligated by Ku70/80-dependent processing. The  HR is expected to increase and saturate at very high complexity.  The alt-NHEJ and SSA are expected to gain weight when canonical repair is compromised.

For overall probability, the class-specific probabilities should be combined using the initial lesion fractions:
\begin{displaymath}
p_{j}^{\mathrm{eff}}= \frac{S_{0}p_{j,S}+C_{0}p_{j,C}}{S_{0}+C_{0}}= (1-f_{c})p_{j,S}+f_{c}p_{j,C}.
\end{displaymath}
Thus, LET and GNP concentration affect the effective pathway allocation partly by increasing \(f_c\), even if the class-specific pathway probabilities remain fixed.

\subsection{Correct repair, misrepair and residual damage}

Pathway allocation specifies the fraction of recognized lesions
assigned to each repair mechanism. It must therefore be distinguished
from the subsequent rate of pathway processing. Let $k_{j,x}$ denote
the processing-rate coefficient of pathway $j$ for lesion class $x$.
The corresponding processing flux is
\begin{equation}
J_{j,x}(t)
=
g_{\mathrm{rec}}[P(t)]\,
p_{j,x}(t)\,
k_{j,x}X_x(t),
\label{eq:pathway-flux}
\end{equation}
where
\begin{equation}
X_S(t)=S(t),
\qquad
X_C(t)=C(t),
\label{eq:X-lesion-class}
\end{equation}
and the recognition gate is
\begin{equation}
g_{\mathrm{rec}}(P)
=
\frac{P^{h_P}}
{K_P^{h_P}+P^{h_P}}.
\label{eq:recognition-gate-allocation}
\end{equation}
The recognition gate controls whether the recruited machinery is
sufficient for processing, $p_{j,x}$ controls pathway selection, and
$k_{j,x}$ controls the rate after selection.

The resulting simple- and complex-DSB equations are (see appendix \ref{sec:Solutions} for details)
\begin{subequations}
\label{eq:coupled-SC-allocation}
\begin{align}
\frac{dS}{dt}
={}&
-k_{SC}S
-
g_{\mathrm{rec}}[P(t)]
\sum_{j\in\{N,H,A\}}
p_{j,S}(t)k_{j,S}S,
\label{EqS}
\\
\frac{dC}{dt}
={}&
k_{SC}S
-
g_{\mathrm{rec}}[P(t)]
\sum_{j\in\{N,H,A\}}
p_{j,C}(t)k_{j,C}C.
\label{EqC}
\end{align}
\end{subequations}

For a reduced form, assume that over the relevant repair interval,
\begin{displaymath}
g_{\mathrm{rec}}[P(t)] \sim  \bar{g}_{\mathrm{rec}}, \qquad p_{j,x}\sim \bar{p}_{j,x},\qquad \mu_{j,x}=\mathrm{constant}
\end{displaymath}
and define
\begin{eqnarray}
a_{S} &=& {\bar g}_{\mathrm{rec}}\sum_{j}{\bar p}_{j,S}k_{j,S}, 
\qquad a_{C}= {\bar g}_{\mathrm{rec}}\sum_{j}{\bar p}_{j,C}k_{j,C}, \nonumber\\
\lambda_{S} & =&  k_{SC}+a_{S}, \qquad  \lambda_{C}= a_{C},\nonumber
\end{eqnarray}
the constant-coefficient analytical solutions for simple and complex DSB numbers are given by
\begin{eqnarray}
S(t) & = & S_{0}e^{-\lambda_{S}t}, \nonumber\\
C(t) & = & C_{0}e^{-\lambda_{C}t}+ \frac{k_{SC}S_{0}}{\lambda_{C}-\lambda_{S}}\bigg(e^{-\lambda_{S}t} -e^{-\lambda_{C}t} \bigg)\nonumber
\end{eqnarray}
The time-integrated simple and complex DSB populations are
\begin{eqnarray}
\mathcal{I}_{S}(t) & = & \int_{0}^{t}S(s)ds = \frac{S_{0}}{\lambda_{S}}(1-e^{-\lambda_{S}t})\\
\mathcal{I}_{C}(t) & = &  \int_{0}^{t}C(s)ds = \frac{S_{0}}{\lambda_{C}}(1-e^{-\lambda_{C}t}) \nonumber\\
& & + \frac{k_{SC}S_{0}}{\lambda_{C}-\lambda_{S}}
\left[\frac{ 1-e^{-\lambda_{S}t}}{\lambda_{S}}- \frac{ 1-e^{-\lambda_{C}t}}{\lambda_{C}}\right]
\end{eqnarray}

Let $\mu_{j,x}$ denote the conditional fidelity of pathway $j$ for a
lesion of class $x$. Thus, $\mu_{j,x}$ is the probability that pathway
processing restores the original or a functionally equivalent DNA
sequence without producing a biologically significant rearrangement.
The complementary probability $1-\mu_{j,x}$ represents misrepair.

Because simple, isolated breaks are structurally stable and undergo minimal alteration across radiation tracking fields, their fidelity drops linearly and slowly with track density and nanoparticle concentration \cite{Stewart2001,Sakata2020}
\begin{equation}
\mu_{S}(L,C_{GNP}) = \mu_{S,0}-\kappa_{S} L(1+\gamma_{S}C_{GNP}).
\end{equation}
The complex and regional multi-break clusters sustain massive local crowding from secondary low-energy Auger cascades; their probability of error-free religation scales down exponentially as a function of local complexity:
\begin{equation}
\mu_{C}(L,C_{GNP}) = \mu_{C,0}{ \mathrm{exp}}\left[-\beta_{C}L(1+\gamma_{S}C_{GNP})\right].
\end{equation}
where $\mu_{S:C,0}$ and $\mu_{C,0}$ are the baseline simple ligation and complex ligation probabilities, $\kappa_{S}$ and $\kappa_{C}$ are linear simple breakdown and exponential complex attenuation coefficients, and $\gamma_{S,C}$ are GNP mass scaling multipliers for simple lesions and complex clusters, respectively.  The baseline scaling parameters and initial states are defined in Table  \ref{tab:Fedilities}.

\begin{table*}[h]
\caption {Baseline scaling parameters and initial states \cite{Lampe2017,Hosseini2022}. }
\label{tab:Fedilities}
\begin{ruledtabular}
\begin{tabular}{cccccc}
Baseline intrinsic  & Track-structure & Nanoparticle Auger \\
efficiencies ($\%)$  & sensitivity factors ($\mu$m/keV)  & amplification multipliers (Con./unit) \\ \hline
$\mu_{S,0}=98.5$ & $\kappa_{S}= 1.2\times 10^{-4}$  & $\gamma_{S}=0.015$ \\
$\mu_{C,0}=88.0$ & $\beta_{C}= 2.8\times 10^{-2}$ & $\gamma_{C}=0.048$ \\  
\end{tabular}
\end{ruledtabular}
\end{table*}

We define the class-specific effective correct-repair rate coefficients
\begin{subequations}
\label{eq:correct-rate-coefficients}
\begin{align}
k_{c,S} ={}& \sum_{j\in\{N,H,A\}} \mu_{j,S}p_{j,S}k_{j,S},\\
k_{c,C}={}&\sum_{j\in\{N,H,A\}}\mu_{j,C}p_{j,C}k_{j,C}.
\end{align}
\end{subequations}
where \(k_{j,S}\) and \(k_{j,C}\) are the repair rate constants through pathway \(j\) for simple (\(S\)) and complex (\(C\)) DNA double-strand breaks, respectively. The  baseline kinetic constants (\(k_{j,S}\) and \(k_{j,C}\)) are given in Table \ref{tab:RPConstants}. 
\begin{table*}[h]
\caption {Intrinsic DNA fixed baseline  repair pathway kinetic constants \cite{Stewart2001,Lampe2017,Incerti2022}.}
\label{tab:RPConstants}
\begin{ruledtabular}
\begin{tabular}{cccccc}
Repair channel &   Intrinsic simple rate   & Intrinsic complex rate\\
                         &   $k_{j,S}$ ($h^{-1}$)  & $k_{j,C}$ ($h^{-1}$)\\ \hline
c-NHEJ ($N$) &   0.88  & 0.12\\
HR ($H$) & 0.15 & 0.065\\
Alt-EJ ($A$) & 0.05 & 0.022\\ 
\end{tabular}
\end{ruledtabular}
\end{table*}

The cumulative number of correctly repaired DSBs, $R_c(t)$, then
satisfies
\begin{equation}
\frac{dR_c}{dt}
=
k_{c,S}S(t)+k_{c,C}C(t).
\label{eq:correct-repair-rate}
\end{equation}
Equivalently,
\begin{equation}
\frac{dR_c}{dt}
=
\sum_{j\in\{N,H,A\}}
\left[
\mu_{j,S}p_{j,S}k_{j,S}S(t)
+
\mu_{j,C}p_{j,C}k_{j,C}C(t)
\right].
\label{eq:correct-repair-expanded}
\end{equation}
With $R_c(0)=0$, the cumulative correct-repair population is
\begin{equation}
R_c(t)
=
k_{c,S}\mathcal I_S(t)
+
k_{c,C}\mathcal I_C(t),
\label{eq:correct-repair-integrated}
\end{equation}
where
\begin{equation}
\mathcal I_S(t)=\int_0^tS(t')\,dt',
\qquad
\mathcal I_C(t)=\int_0^tC(t')\,dt'.
\label{eq:integrated-SC}
\end{equation}

For $\lambda_C\ne\lambda_S$, the required time integrals are
\begin{equation}
\mathcal I_S(t)
=
\frac{S_0}{\lambda_S}
\left(1-e^{-\lambda_St}\right),
\label{eq:IS}
\end{equation}
and
\begin{align}
\mathcal I_C(t)
={}&
\frac{C_0}{\lambda_C}
\left(1-e^{-\lambda_Ct}\right)
\nonumber\\
&+
\frac{k_{SC}S_0}{\lambda_C-\lambda_S}
\left[
\frac{1-e^{-\lambda_St}}{\lambda_S}
-
\frac{1-e^{-\lambda_Ct}}{\lambda_C}
\right].
\label{eq:IC}
\end{align}

Misrepair includes -end joining, sequence loss, incomplete
restoration, translocation, inversion, dicentric formation, and other
structurally altered outcomes. Define the class-specific misrepair
coefficients
\begin{subequations}
\label{eq:misrepair-rate-coefficients}
\begin{align}
k_{m,S}
={}&
\sum_{j\in\{N,H,A\}}
(1-\mu_{j,S})p_{j,S}k_{j,S},
\\
k_{m,C}
={}&
\sum_{j\in\{N,H,A\}}
(1-\mu_{j,C})p_{j,C}k_{j,C}.
\end{align}
\end{subequations}
The cumulative misrepair population satisfies
\begin{equation}
\frac{dR_m}{dt}
=
k_{m,S}S(t)+k_{m,C}C(t).
\label{eq:misrepair-rate}
\end{equation}
Equivalently,
\begin{eqnarray}
\frac{dR_m}{dt} & = & \sum_{j\in\{N,H,A\}} \left[
(1-\mu_{j,S})p_{j,S}k_{j,S}S(t) \right. \nonumber\\
& & \left.
+
(1-\mu_{j,C})p_{j,C}k_{j,C}C(t)
\right].
\label{eq:misrepair-expanded}
\end{eqnarray}
For $R_m(0)=0$, integration gives
\begin{equation}
R_m(t)
=
k_{m,S}\mathcal I_S(t)
+
k_{m,C}\mathcal I_C(t).
\label{eq:misrepair-integrated}
\end{equation}


The residual DSBs per necleus at time $t$ is defined as
\begin{equation}
U(t)=S(t)+C(t).
\label{eq:U-residual}
\end{equation}
Using the simple- and complex-lesion solutions gives
\begin{align}
U(t)
={}&
S_0e^{-\lambda_St}
+
C_0e^{-\lambda_Ct}
\nonumber\\
&+
\frac{k_{SC}S_0}{\lambda_C-\lambda_S}
\left(
e^{-\lambda_St}-e^{-\lambda_Ct}
\right).
\label{eq:U-explicit}
\end{align}
Addition of Eqs. (\ref{EqS}) and \ref{EqC}) gives
\begin{equation}
\frac{dU}{dt}
=
-k_{S,\mathrm{rep}}S
-k_{C,\mathrm{rep}}C,
\label{eq:U-rate}
\end{equation}
where
\begin{equation}
k_{x,\mathrm{rep}}
=
\sum_{j\in\{N,H,A\}}p_{j,x}k_{j,x}.
\end{equation}
The absence of $k_{SC}$ from Eq.~(\ref{eq:U-rate}) shows that
simple-to-complex conversion redistributes unrepaired DSBs but does not
change their total number.

For comparison with an experimental residual-focus assay, the
biologically relevant residual lesions per nucleus are evaluated at the assay time
$t_b$,
\begin{equation}
U_b\equiv U(t_b).
\label{eq:Ub}
\end{equation}
The time $t_b$ must therefore be reported together with the residual damage prediction. In the absence of additional DSB production, enzymatic loss channels, or spontaneous lesion disappearance, the total lesion balance is
\begin{equation}
N_0 = S(t)+C(t)+R_c(t)+R_m(t).
\label{eq:lesion-conservation}
\end{equation}

\section{Monte Carlo Track Structure Simulation}

The Geant4-DNA Monte Carlo track structure simulation toolkit \cite{Schuemann2019,Incerti2010Project,Incerti2010Comparison} was utilised to model the radiobiological response of asynchronous human fibroblast cells (AG1522, 6.4~Gbp genome) to proton irradiation across a therapeutically relevant LET range of 10 to 50~keV/$\mu$m \cite{Bernal2015Review,Incerti2018,Shin2021}. The simulation incorporates the physical, physico-chemical, and chemical stages of radiation action at the nanometer scale, enabling quantification of initial double-strand break (DSB) payloads, complex lesion configurations, and their subsequent processing through competing DNA repair pathways. A key innovation of this modelling framework is the integration of 15~nm GNPs at varying concentrations (5$\times$, 10$\times$, 15$\times$, and 20$\times$ relative to the baseline). These nanoparticles serve as radiosensitising agents by facilitating localised Auger electron emission and enhancing dose deposition within critical cellular targets.

The AG1522 cell line was selected for its well-characterised radiosensitivity, normal human fibroblast phenotype, and diploid genome status, making it an ideal model for investigating radiation-induced DNA damage and repair fidelity. The simulation geometry explicitly represents the cell nucleus as the primary target volume, with DSB induction calculated per nucleus per unit dose. The 6.4~Gbp genome size allows for absolute quantification of lesion numbers, enabling direct comparison with experimental $\gamma$-H2AX foci and pulsed-field gel electrophoresis data. The spatial coordinates and structural complexity of all radiation-induced single-strand breaks (SSBs) and double-strand breaks (DSBs) were scored and exported in the Standard for DNA Damage Simulation (SDDS) format \cite{Schuemann2019}. 

\subsection{Initial DSB Induction and Damage Complexity}

Geant4-DNA simulations reveal a pronounced LET dependence in both the quantity and quality of initial DSB damage (Table~\ref{tab:initial_dsb}). At the lowest LET examined (10~keV/$\mu$m, corresponding to the Bragg peak entrance region), the total DSB yield at 1.5~Gy in the absence of GNPs is $93.6 \pm 2.7$ per cell nucleus. As LET increases to 50~keV/$\mu$m (overkill and saturation interface), the total DSB yield at the same dose more than doubles to $194.4 \pm 4.2$, with a dramatic shift in lesion complexity: simple DSBs decrease to $48.0 \pm 1.6$, while complex and clustered lesions increase to $91.2 \pm 3.1$ and $55.2 \pm 2.4$, respectively.

The shift from simple to complex damage corresponds to an increase in ionization density along the proton track. At higher LET, individual proton tracks deposit energy in more spatially concentrated clusters, resulting in multiply damaged sites (MDS) that are more challenging to repair. The ratio of complex-to-simple double-strand breaks (DSBs) rises from approximately 0.46 at 10~keV/$\mu$m to 1.90 at 50~keV/$\mu$m, quantitatively illustrating the shift in damage spectrum as LET increases.

\begin{table*}[h]
\caption{Initial DSB payloads per cell nucleus (AG1522, 6.4~Gbp) at 1.5~Gy as a function of LET (0$\times$ GNP control baseline). Values are mean $\pm$ standard deviation.}
\label{tab:initial_dsb}
\begin{ruledtabular}
\begin{tabular}{cccccc}
LET (keV/$\mu$m) & Simple & Complex & Clustered & Total \\
\hline
10 & $57.6 \pm 2.4$ & $26.4 \pm 1.1$ & $9.6 \pm 0.6$ & $93.6 \pm 2.7$ \\
20 & $62.4 \pm 2.2$ & $45.6 \pm 1.8$ & $21.6 \pm 1.1$ & $129.6 \pm 3.1$ \\
30 & $62.4 \pm 2.1$ & $57.6 \pm 2.2$ & $33.6 \pm 1.6$ & $153.6 \pm 3.4$ \\
40 & $57.6 \pm 1.9$ & $79.2 \pm 2.8$ & $45.6 \pm 2.1$ & $182.4 \pm 3.9$ \\
50 & $48.0 \pm 1.6$ & $91.2 \pm 3.1$ & $55.2 \pm 2.4$ & $194.4 \pm 4.2$ \\
\end{tabular}
\end{ruledtabular}
\end{table*}


The incorporation of 15~nm GNPs at increasing concentrations systematically amplifies DSB induction across all LET values (Table~\ref{tab:gnp_dsb}). Notably, the proportional increase is most pronounced for complex and clustered lesions: at 20$\times$ GNP concentration, complex DSBs increase by 70.8\% (from 26.4 to 45.1) and clustered DSBs by 132.3\% (from 9.6 to 22.3), compared to 57.6 to 56.4 ($-2.1\%$) for simple DSBs. This differential amplification results from the physical properties of Auger electron emission from gold nanoparticles (GNPs). When gold atoms interact with primary proton radiation or secondary electrons, they emit cascades of low-energy Auger electrons (typically 10 to 100 eV) that deposit energy within highly localised nanometer-scale volumes surrounding the GNP. These localised energy deposits preferentially induce clustered damage sites rather than isolated, simple breaks, thereby shifting the damage spectrum toward greater complexity.

\begin{table*}[h]
\caption{Initial DSB payloads per cell nucleus at 1.5~Gy as a function of LET and GNP concentration (15~nm GNPs). Values are mean $\pm$ standard deviation.}
\label{tab:gnp_dsb}
\begin{ruledtabular}
\begin{tabular}{cccccc}
LET (keV/$\mu$m) & GNP Conc. & Simple & Complex & Clustered & Total \\
\hline
10 & 5$\times$  & $59.3 \pm 2.1$ & $31.0 \pm 1.2$ & $12.5 \pm 0.6$ & $102.8 \pm 2.5$ \\
10 & 10$\times$ & $60.0 \pm 2.2$ & $36.0 \pm 1.4$ & $15.7 \pm 0.8$ & $111.7 \pm 2.7$ \\
10 & 15$\times$ & $58.7 \pm 2.3$ & $40.8 \pm 1.7$ & $18.9 \pm 1.0$ & $118.4 \pm 3.0$ \\
10 & 20$\times$ & $56.4 \pm 2.4$ & $45.1 \pm 1.9$ & $22.3 \pm 1.2$ & $123.8 \pm 3.3$ \\
20 & 5$\times$  & $64.3 \pm 2.1$ & $53.4 \pm 1.9$ & $28.1 \pm 1.3$ & $145.8 \pm 3.1$ \\
20 & 10$\times$ & $65.0 \pm 2.2$ & $62.0 \pm 2.3$ & $35.3 \pm 1.6$ & $162.3 \pm 3.6$ \\
20 & 15$\times$ & $63.6 \pm 2.3$ & $70.2 \pm 2.7$ & $42.5 \pm 2.0$ & $176.3 \pm 4.1$ \\
20 & 20$\times$ & $61.1 \pm 2.4$ & $77.6 \pm 3.0$ & $50.1 \pm 2.4$ & $188.8 \pm 4.5$ \\
30 & 5$\times$  & $64.3 \pm 2.0$ & $67.4 \pm 2.3$ & $43.7 \pm 1.8$ & $175.4 \pm 3.4$ \\
30 & 10$\times$ & $65.0 \pm 2.1$ & $78.4 \pm 2.7$ & $54.8 \pm 2.3$ & $198.2 \pm 4.1$ \\
30 & 15$\times$ & $63.6 \pm 2.3$ & $88.7 \pm 3.2$ & $66.2 \pm 2.8$ & $218.5 \pm 4.8$ \\
30 & 20$\times$ & $61.1 \pm 2.4$ & $98.1 \pm 3.6$ & $78.0 \pm 3.4$ & $237.2 \pm 5.5$ \\
40 & 5$\times$  & $59.3 \pm 1.8$ & $92.4 \pm 3.3$ & $59.3 \pm 2.6$ & $211.0 \pm 4.6$ \\
40 & 10$\times$ & $60.0 \pm 2.0$ & $107.5 \pm 3.9$ & $74.4 \pm 3.3$ & $241.9 \pm 5.5$ \\
40 & 15$\times$ & $58.7 \pm 2.1$ & $121.7 \pm 4.6$ & $89.8 \pm 4.1$ & $270.2 \pm 6.5$ \\
40 & 20$\times$ & $56.4 \pm 2.2$ & $134.4 \pm 5.3$ & $105.8 \pm 5.0$ & $296.6 \pm 7.6$ \\
50 & 5$\times$  & $49.5 \pm 1.5$ & $106.5 \pm 3.6$ & $64.4 \pm 2.8$ & $220.4 \pm 4.8$ \\
50 & 10$\times$ & $50.1 \pm 1.6$ & $124.0 \pm 4.3$ & $80.8 \pm 3.5$ & $254.9 \pm 5.7$ \\
50 & 15$\times$ & $49.0 \pm 1.8$ & $140.4 \pm 5.0$ & $97.4 \pm 4.3$ & $286.8 \pm 6.8$ \\
50 & 20$\times$ & $47.1 \pm 1.9$ & $155.0 \pm 5.7$ & $114.7 \pm 5.2$ & $316.8 \pm 8.0$ \\
\end{tabular}
\end{ruledtabular}
\end{table*}
The enhancement mediated by GNPs is dependent on LET, with the largest absolute increases occurring at intermediate LET values (30 to 40 keV/$\mu$m), where the spatial and temporal overlap between the primary proton track and Auger electron cascades is optimal. At 40 keV/$\mu$m and a dose of 1.5 Gy, the total number of double-strand breaks (DSBs) increases from $182.4 \pm 3.9$ (0$\times$) to $296.6 \pm 7.6$ (20$\times$), representing a 62.6\% increase. This dependence on LET demonstrates the interaction between direct proton-induced damage and secondary electron effects mediated by GNPs.

\subsection{DNA Repair Pathway Allocation and Fidelity}
\subsubsection{Cell Cycle-Dependent Pathway Selection}
The simulation incorporates cell cycle-dependent DNA repair pathway allocation probabilities, recognizing that AG1522 cells irradiated asynchronously present a heterogeneous population with respect to repair capacity (Tables~\ref{tab:g1_pathway} and~\ref{tab:g2_pathway}).
The probabilities are applied to both the simple- and complex-DSB processing fluxes unless class-dependent allocations are introduced explicitly. In G1/early-S phase cells, the primary repair pathways are canonical non-homologous end joining (c-NHEJ), homologous recombination (HR), and alternative NHEJ (Alt-NHEJ). At 10~keV/$\mu$m and 5$\times$ GNP concentration, the allocation is 66\% c-NHEJ, 22\% HR, and 12\% Alt-NHEJ. As LET increases to 50~keV/$\mu$m, c-NHEJ allocation decreases to 11\% (20$\times$ GNP), while HR and Alt-NHEJ increase to 51\% and 38\%, respectively.
\begin{table*}[h]
\caption{G1/Early-S phase pathway allocation probabilities (\%) as a function of LET and GNP concentration. }
\label{tab:g1_pathway}
\begin{ruledtabular}
\begin{tabular}{ccccccccc}
LET (keV/$\mu$m) & GNP Conc. & c-NHEJ & HR & Alt-NHEJ & Primary Molecular Routing Driver \\
\hline
10 & 5$\times$  & 66\% & 22\% & 12\% & Isolated Auger electrons cause minor flanking base lesions. \\
10 & 10$\times$ & 62\% & 24\% & 14\% & \\
10 & 15$\times$ & 58\% & 26\% & 16\% & \\
10 & 20$\times$ & 54\% & 28\% & 18\% & Dense local groupings begin to inhibit Ku70/80 binding. \\
20 & 5$\times$  & 52\% & 30\% & 18\% & Complex configurations outpace simple breaks. \\
20 & 10$\times$ & 47\% & 33\% & 20\% & \\
20 & 15$\times$ & 42\% & 35\% & 23\% & Alt-NHEJ matches c-NHEJ processing frequency. \\
20 & 20$\times$ & 37\% & 37\% & 26\% & Core Bragg peak entry shifts dominant mechanisms. \\
30 & 10$\times$ & 35\% & 40\% & 25\% & \\
30 & 15$\times$ & 30\% & 42\% & 28\% & Severely complex ends dominate the damage payload. \\
30 & 20$\times$ & 25\% & 44\% & 31\% & Peak theoretical track overlap with 15~nm geometries. \\
40 & 10$\times$ & 25\% & 45\% & 30\% & \\
40 & 15$\times$ & 21\% & 47\% & 32\% & c-NHEJ is heavily suppressed by structural clutter. \\
40 & 20$\times$ & 16\% & 49\% & 35\% & Overkill phase; multiple hits concentrate on single loops. \\
50 & 10$\times$ & 18\% & 49\% & 33\% & \\
50 & 15$\times$ & 14\% & 50\% & 36\% & \\
50 & 20$\times$ & 11\% & 51\% & 38\% & Alt-NHEJ serves as the primary end-joining backup. \\
\end{tabular}
\end{ruledtabular}
\end{table*}
\begin{table*}[h]
\caption{Late-S/G2 phase pathway allocation probabilities (\%) as a function of LET and GNP concentration.}
\label{tab:g2_pathway}
\begin{ruledtabular}
\begin{tabular}{cccccccc}
LET (keV/$\mu$m) & GNP Conc. & HR & c-NHEJ & Alt-NHEJ & SSA & Evolutionary Failure \\
\hline
10 & 5$\times$  & 50\% & 38\% & 8\%  & 4\%  & HR functions efficiently on single Auger tracks. \\
10 & 10$\times$ & 51\% & 34\% & 9\%  & 6\%  & \\
10 & 15$\times$ & 52\% & 30\% & 11\% & 7\%  & \\
10 & 20$\times$ & 52\% & 27\% & 12\% & 9\%  & Template-directed loops begin showing edge cuts. \\
20 & 5$\times$  & 51\% & 29\% & 12\% & 8\%  & End resection pathways scale up dynamically. \\
20 & 10$\times$ & 49\% & 25\% & 14\% & 12\% & \\
20 & 15$\times$ & 46\% & 22\% & 17\% & 15\% & \\
20 & 20$\times$ & 43\% & 19\% & 20\% & 18\% & Resection exposes extensive repeat segments. \\
30 & 5$\times$  & 45\% & 23\% & 18\% & 14\% & HR capacity peaks and begins to saturate. \\
30 & 10$\times$ & 41\% & 20\% & 21\% & 18\% & \\
30 & 15$\times$ & 37\% & 17\% & 23\% & 23\% & \\
30 & 20$\times$ & 33\% & 14\% & 26\% & 27\% & \\
40 & 5$\times$  & 39\% & 18\% & 23\% & 20\% & \\
40 & 10$\times$ & 34\% & 15\% & 26\% & 25\% & \\
40 & 15$\times$ & 30\% & 12\% & 28\% & 30\% & \\
40 & 20$\times$ & 26\% & 9\%  & 31\% & 34\% & \\
50 & 5$\times$  & 32\% & 14\% & 27\% & 27\% & \\
50 & 10$\times$ & 28\% & 11\% & 29\% & 32\% & \\
50 & 15$\times$ & 24\% & 9\%  & 31\% & 36\% & \\
50 & 20$\times$ & 20\% & 6\%  & 34\% & 40\% & \\
\end{tabular}
\end{ruledtabular}
\end{table*}

This shift reflects the underlying structural complexity of DNA damage. Simple DSBs with clean ends are efficiently repaired by c-NHEJ, which requires minimal end processing. In contrast, complex and clustered lesions necessitate extensive end resection and homology-directed repair. When HR capacity is exceeded, cells increasingly rely on error-prone Alt-NHEJ.  Therefore, the increase in complex damage associated with higher GNP concentrations directly influences repair pathway selection, shifting the balance from accurate c-NHEJ toward less precise mechanisms.\\

In late-S/G2 phase cells, HR serves as the predominant repair pathway at low LET, accounting for 50–52$\%$ of repair events across GNP concentrations at 10 keV/$\mu$m. This predominance corresponds to the presence of sister chromatids, which provide homologous templates. As LET increases to 50 keV/$\mu$m, the allocation to HR decreases to 20–32$\%$, depending on GNP concentration, while c-NHEJ, Alt-NHEJ, and SSA become increasingly prominent. The marked rise in SSA usage (up to 40$\%$ at 50 keV/$\mu$m, 20$\times$ GNP) indicates extensive end resection that exposes repetitive genomic sequences, thereby facilitating error-prone repair and elevating the risk of genomic rearrangements.

\subsubsection{Repair Fidelity Matrix}

The DNA-repair fidelity and misrepair matrix (Table ~\ref{tab:independent-fidelity-misrepair}) for AG1522 cells
predicts a progressive deterioration in repair accuracy with increasing
proton LET and GNP concentration. The fidelity for simple DSBs,
$\mu_S$, remains comparatively high, decreasing moderately from
$95.5\%$ at $L=10~\mathrm{keV}\,\mu\mathrm{m}^{-1}$ and a
$5\times$ GNP concentration to $91.2\%$ at
$L=50~\mathrm{keV}\,\mu\mathrm{m}^{-1}$ and $20\times$. In contrast,
the fidelity for complex or clustered DSBs, $\mu_C$, decreases
substantially from $70.8\%$ to $8.8\%$ over the same parameter range,
reflecting the increasing difficulty of correctly processing
spatially clustered lesions and chemically damaged DNA termini.
Consequently, the processed-lesion-weighted mean misrepair fraction,
$\overline{f}_{m}=1-\overline{\mu}$, increases from $14.2\%$ to
$70.0\%$. These estimates were obtained using a leave-one-LET-out
validation procedure in which the complete set of concentrations and
doses at the LET being evaluated was excluded from parameter
estimation. The results therefore indicate that the LET- and
GNP-dependent reduction in overall repair fidelity is governed
predominantly by the increasing abundance and reduced fidelity of
complex DSBs, rather than by a substantial loss of simple-DSB repair
fidelity.
\begin{table*}[h]
\centering
\caption{%
Predicted DNA-repair fidelity and misrepair matrix for
AG1522 cells containing 15-nm GNPs. For each LET, the complete block
of concentrations and doses at that LET was excluded from parameter
estimation before prediction. The mean misrepair fraction is
the processed-lesion-weighted quantity
$\overline{f}_{m}=1-\overline{\mu}$.  Values are the model predictions
$\pm$ propagated $1\sigma$ parameter uncertainties.%
}
\label{tab:independent-fidelity-misrepair}
\setlength{\tabcolsep}{7pt}
\renewcommand{\arraystretch}{1.12}
\begin{tabular}{ccccc}
\toprule
LET &
GNP concentration &
Simple-DSB fidelity &
Complex/clustered fidelity &
Mean misrepair fraction
\\
$(\mathrm{keV}\,\mu\mathrm{m}^{-1})$ &
&
$\mu_S$ (\%) &
$\mu_C$ (\%) &
$\overline{f}_{m}$ (\%)
\\
\midrule
10 & $5\times$  & $95.5\pm1.2$ & $70.8\pm3.7$ & $14.2\pm1.6$ \\
10 & $10\times$ & $95.2\pm0.9$ & $66.8\pm2.7$ & $16.9\pm1.2$ \\
10 & $15\times$ & $94.9\pm1.1$ & $62.5\pm3.2$ & $20.2\pm1.6$ \\
10 & $20\times$ & $94.5\pm1.9$ & $58.0\pm5.1$ & $23.9\pm2.7$ \\
\addlinespace

20 & $5\times$  & $95.5\pm0.7$ & $58.9\pm2.2$ & $23.0\pm1.2$ \\
20 & $10\times$ & $94.9\pm0.5$ & $52.5\pm1.5$ & $27.9\pm0.8$ \\
20 & $15\times$ & $94.3\pm0.7$ & $46.0\pm1.7$ & $33.3\pm1.0$ \\
20 & $20\times$ & $93.6\pm1.2$ & $39.7\pm2.5$ & $39.0\pm1.6$ \\
\addlinespace

30 & $5\times$  & $94.6\pm0.6$ & $44.5\pm1.8$ & $33.7\pm1.1$ \\
30 & $10\times$ & $94.1\pm0.4$ & $37.6\pm1.2$ & $39.6\pm0.7$ \\
30 & $15\times$ & $93.6\pm0.5$ & $31.2\pm1.2$ & $45.5\pm0.8$ \\
30 & $20\times$ & $93.0\pm0.9$ & $25.4\pm1.7$ & $51.1\pm1.2$ \\
\addlinespace

40 & $5\times$  & $93.6\pm1.0$ & $30.7\pm2.3$ & $46.7\pm1.5$ \\
40 & $10\times$ & $93.1\pm0.7$ & $24.6\pm1.3$ & $52.6\pm0.9$ \\
40 & $15\times$ & $92.6\pm0.8$ & $19.3\pm1.3$ & $58.0\pm0.9$ \\
40 & $20\times$ & $92.1\pm1.3$ & $15.0\pm1.6$ & $62.8\pm1.2$ \\
\addlinespace

50 & $5\times$  & $92.3\pm2.4$ & $21.0\pm3.2$ & $56.8\pm2.4$ \\
50 & $10\times$ & $91.9\pm1.6$ & $15.9\pm1.7$ & $61.9\pm1.3$ \\
50 & $15\times$ & $91.6\pm1.6$ & $11.9\pm1.4$ & $66.4\pm1.2$ \\
50 & $20\times$ & $91.2\pm2.6$ &  $8.8\pm1.7$ & $70.0\pm1.5$ \\
\bottomrule
\end{tabular}
\end{table*}

\subsection{Kinetic Repair-Rate Coefficients}

The kinetic repair-rate coefficients matrix  provides a quantitative description of the time-dependent processing of DSBs. The repair kinetics are best described by a biphasic exponential model with fast and slow components. The fast coefficient ($k_{\mathrm{fast}}$) represents the rapid phase of repair, likely corresponding to simple DSBs and those complex lesions that are efficiently processed by c-NHEJ. The slow coefficient ($k_{\mathrm{slow}}$) represents the persistent fraction of damage, typically complex and clustered lesions requiring HR or Alt-NHEJ, or those that remain unrepaired.
\begin{table*}[h]
\centering
\caption{%
Kinetic repair-rate coefficient matrix for 15-nm GNPs in
AG1522 cells. The  coefficients were predicted using
leave-one-LET-out validation. }
\label{tab:revised-repair-rates}
\setlength{\tabcolsep}{7pt}
\renewcommand{\arraystretch}{1.12}
\begin{tabular}{cccccc}
\toprule
LET &
GNP concentration &
Fast coefficient &
Slow coefficient &
Fast half-life &
Slow half-life
\\
$(\mathrm{keV}\,\mu\mathrm{m}^{-1})$ &
&
$k_S$ ($\mathrm{h}^{-1}$) &
$k_C$ ($\mathrm{h}^{-1}$) &
$t_{1/2,S}$ (min) &
$t_{1/2,C}$ (h)
\\
\midrule
10 & $5\times$  & $0.74\pm0.03$ & $0.0828\pm0.0016$ & 56 &  8.4 \\
10 & $10\times$ & $0.72\pm0.03$ & $0.0803\pm0.0010$ & 58 &  8.6 \\
10 & $15\times$ & $0.70\pm0.04$ & $0.0779\pm0.0010$ & 59 &  8.9 \\
10 & $20\times$ & $0.68\pm0.04$ & $0.0755\pm0.0015$ & 61 &  9.2 \\
\addlinespace

20 & $5\times$  & $0.72\pm0.03$ & $0.0675\pm0.0012$ & 58 & 10.3 \\
20 & $10\times$ & $0.70\pm0.04$ & $0.0629\pm0.0008$ & 59 & 11.0 \\
20 & $15\times$ & $0.67\pm0.04$ & $0.0587\pm0.0007$ & 62 & 11.8 \\
20 & $20\times$ & $0.65\pm0.05$ & $0.0547\pm0.0010$ & 64 & 12.7 \\
\addlinespace

30 & $5\times$  & $0.69\pm0.04$ & $0.0577\pm0.0008$ & 60 & 12.0 \\
30 & $10\times$ & $0.66\pm0.04$ & $0.0528\pm0.0005$ & 63 & 13.1 \\
30 & $15\times$ & $0.64\pm0.04$ & $0.0483\pm0.0004$ & 65 & 14.4 \\
30 & $20\times$ & $0.62\pm0.05$ & $0.0441\pm0.0006$ & 67 & 15.7 \\
\addlinespace

40 & $5\times$  & $0.66\pm0.04$ & $0.0500\pm0.0011$ & 63 & 13.9 \\
40 & $10\times$ & $0.63\pm0.04$ & $0.0449\pm0.0006$ & 66 & 15.4 \\
40 & $15\times$ & $0.61\pm0.05$ & $0.0403\pm0.0006$ & 68 & 17.2 \\
40 & $20\times$ & $0.59\pm0.05$ & $0.0362\pm0.0008$ & 70 & 19.1 \\
\addlinespace

50 & $5\times$  & $0.62\pm0.04$ & $0.0430\pm0.0018$ & 67 & 16.1 \\
50 & $10\times$ & $0.59\pm0.05$ & $0.0378\pm0.0010$ & 70 & 18.3 \\
50 & $15\times$ & $0.57\pm0.05$ & $0.0332\pm0.0009$ & 73 & 20.9 \\
50 & $20\times$ & $0.55\pm0.06$ & $0.0292\pm0.0012$ & 76 & 23.8 \\
\bottomrule
\end{tabular}
\end{table*}

The  matrix for AG1522 cells containing 15-nm GNPs (Table~\ref{tab:revised-repair-rates})  exhibits a systematic reduction in both the fast and slow
repair coefficients with increasing proton LET and GNP concentration.
The simple-DSB repair coefficient, $k_S$, decreases moderately from
$0.74\pm0.03~\mathrm{h}^{-1}$ at
$L=10~\mathrm{keV}\,\mu\mathrm{m}^{-1}$ and a $5\times$ GNP
concentration to $0.55\pm0.06~\mathrm{h}^{-1}$ at
$L=50~\mathrm{keV}\,\mu\mathrm{m}^{-1}$ and $20\times$, corresponding
to an increase in the simple-DSB half-life from approximately
$56$ to $76$ min. A stronger dependence is obtained for the
complex-DSB repair coefficient, $k_C$, which decreases from
$0.0828\pm0.0016~\mathrm{h}^{-1}$ to
$0.0292\pm0.0012~\mathrm{h}^{-1}$ over the same range. The associated
complex-DSB half-life therefore increases from approximately
$8.4$ to $23.8$ h. While the fast-component half-life changes by only ~20 minutes across the entire matrix, the slow-component half-life increases by over 100 hours at the high-LET, high-concentration limits. This quantitatively demonstrates how dense secondary Auger electron clusters physically block standard loop processing. The resulting trends indicate that increasing
LET and GNP concentration have a relatively modest effect on the rapid
processing of simple lesions but substantially delay the resolution of
complex or clustered DSBs, consistent with the greater structural and
chemical complexity of the DNA damage generated under these
conditions. Because the intrinsic repair constants (\(k_{j}\)) are completely fixed (\ref{tab:RPConstants}), the deceleration of the cell's overall clearing capacity is purely an effect of the declining repair fidelities (\(\mu_S, \mu_C\)) shown in Table \ref{tab:independent-fidelity-misrepair}. The dynamic shifts observed in total cell repair rates across different LET and GNP mass steps are driven entirely by how the cell shifts its pathway allocation probabilities (\(p_{j}\)) and changes its repair fidelities (\(\mu _{j}\)) as the physical damage tracking density climbs. This kinetic signature has profound implications for the temporal window of repair and the probability of residual damage at clinically relevant timepoints (e.g., 24~h post-irradiation). The slow component represents a reservoir of damage that persists for extended periods, potentially leading to late effects, genomic instability, or mitotic catastrophe.

\section{Residual DSBs at 24 Hours Post-Irradiation}

The absolute number of residual (unrepaired) DSBs at 24~h post-irradiation (Tables~ \ref{tab:residual-DSB-LET10}--\ref{tab:residual-DSB-LET20}) provides a direct link between initial damage, repair kinetics, and cellular outcome. At 10~keV/$\mu$m and 1.5~Gy, residual DSBs increase from $6.7 \pm 0.3$ (5$\times$ GNP) to $13.6 \pm 0.7$ (20$\times$ GNP). At 50~keV/$\mu$m and 3.5~Gy, residual DSBs range from $144.0 \pm 7.2$ (5$\times$) to $317.9 \pm 15.9$ (20$\times$), representing a 2.2-fold enhancement.\\
\begin{table}[htbp]
    \centering
    \begin{minipage}[t]{0.48\textwidth}
        \centering
\caption{%
Predicted residual DSB lesions per AG1522 cell nucleus
at $24$ h after irradiation for
$L=10~\mathrm{keV}\,\mu\mathrm{m}^{-1}$. }
\label{tab:residual-DSB-LET10}
        \small 
        \begin{tabular}{ccccc} 
            \hline
\toprule
Dose (Gy) &
$5\times$ GNP &
$10\times$ GNP &
$15\times$ GNP &
$20\times$ GNP
\\
\midrule
            1.5 & \(6.0\pm0.2\) & \(7.5\pm0.2\) & \(9.2\pm0.2\) & \(11.0\pm0.4\) \\
            2.0 & \(8.0\pm0.3\) & \(10.0\pm0.3\) & \(12.3\pm0.3\) & \(14.7\pm0.5\) \\
            2.5 & \(10.0\pm0.4\) & \(12.5\pm0.3\) & \(15.3\pm0.4\) & \(18.3\pm0.7\) \\
            3.0 & \(11.9\pm0.5\) & \(15.0\pm0.4\) & \(18.4\pm0.4\) & \(22.0\pm0.8\) \\
            3.5 & \(13.9\pm0.6\) & \(17.6\pm0.4\) & \(21.5\pm0.5\) & \(25.6\pm0.9\) \\
            \hline
        \end{tabular}
    \end{minipage}
    \hfill 
    \begin{minipage}[t]{0.48\textwidth}
        \centering
\caption{%
Predicted residual DSB lesions per AG1522 cell nucleus
at $24$ h after irradiation for
$L=20~\mathrm{keV}\,\mu\mathrm{m}^{-1}$. }
\label{tab:residual-DSB-LET20}
        \small
        \begin{tabular}{ccccc} 
            \hline
\toprule
Dose (Gy) &
$5\times$ GNP &
$10\times$ GNP &
$15\times$ GNP &
$20\times$ GNP
\\
\midrule
1.5 & $16.1\pm0.5$ & $21.5\pm0.4$ & $27.6\pm0.5$ & $34.4\pm0.8$ \\
2.0 & $21.5\pm0.6$ & $28.7\pm0.5$ & $36.8\pm0.6$ & $45.8\pm1.1$ \\
2.5 & $26.9\pm0.8$ & $35.8\pm0.7$ & $46.0\pm0.8$ & $57.3\pm1.4$ \\
3.0 & $32.3\pm1.0$ & $43.0\pm0.8$ & $55.2\pm0.9$ & $68.7\pm1.7$ \\
3.5 & $37.6\pm1.1$ & $50.1\pm0.9$ & $64.4\pm1.1$ & $80.2\pm1.9$ \\
            \hline
        \end{tabular}
    \end{minipage}
\end{table}
\begin{figure*}[h]
\centering
\includegraphics[width=0.50\textwidth]
{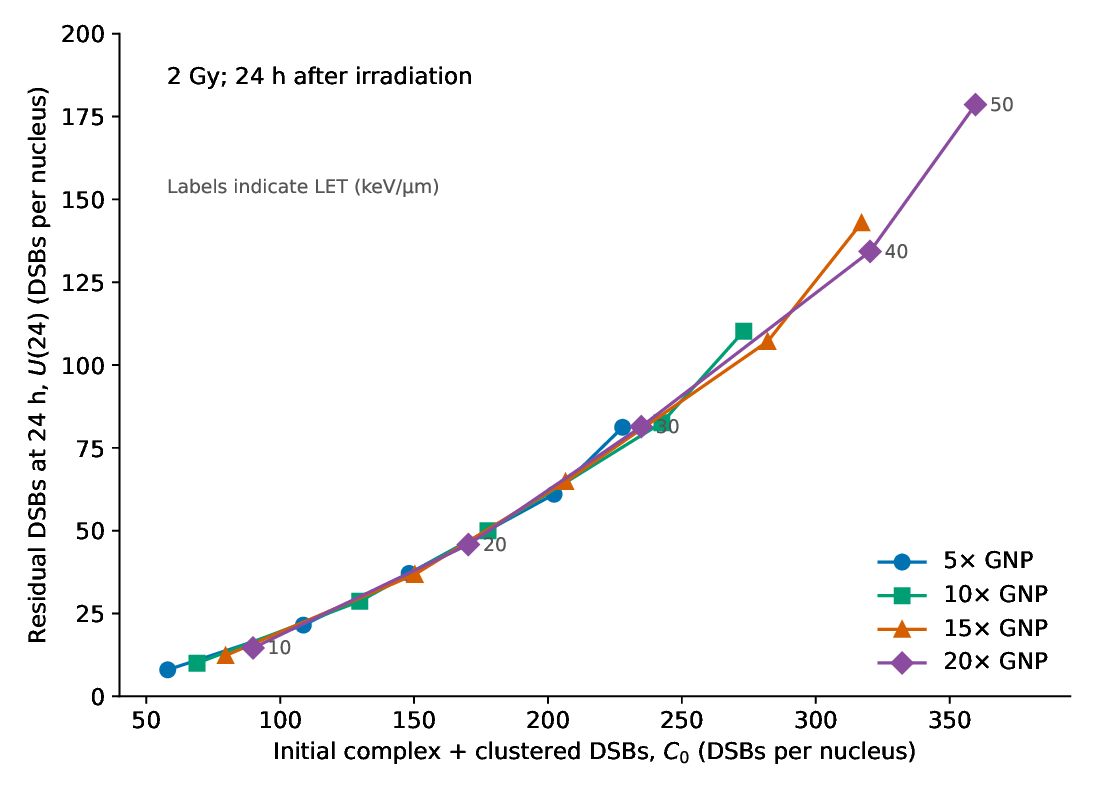}
\caption{%
Predicted residual DSBs at 24 h versus the initial complex/clustered DSB population, \(C_0=N_{\mathrm{complex}}+N_{\mathrm{clustered}}\), in AG1522 cells following 2 Gy proton irradiation.  Colours and symbols indicate relative GNP concentrations; lines connect successive LET conditions from 10 to 50 keV µm\(^{-1}\). Numerical labels identify LET values along the \(20\times\) GNP series.}
\label{fig:Residual_DSB_vs_CC}
\end{figure*}

The residual DSBs per nucleus  exhibit a supra-linear dependence on both LET and GNP concentration, reflecting the combined effects of increased initial damage and decreased repair fidelity. Plotting residual DSBs against initial complex/clustered DSBs reveals a strong correlation ($R^2 > 0.95$), suggesting that the complex lesion fraction is the primary determinant of persistent damage. This relationship provides a mechanistic basis for the enhanced relative biological effectiveness (RBE) observed with increasing LET and GNP concentration.

\subsection{Comparison at 2 Gy}
Table~\ref{tab:endpoint-validation-2Gy}  compares the independently predicted repair endpoints
with the corresponding Geant4-DNA-derived values at
$D_p=2~\mathrm{Gy}$. Among the three endpoints, residual damage at
24~h exhibits the strongest quantitative agreement, with
$\mathrm{RMSE}=3.14$ and $\mathrm{MAE}=2.85$ DSB lesions per nucleus.
The errors are appreciably larger for correct repair
($\mathrm{RMSE}=20.53$, $\mathrm{MAE}=17.88$) and misrepair
($\mathrm{RMSE}=19.81$, $\mathrm{MAE}=17.37$). The discrepancy in
these repair-fate endpoints generally increases with LET and GNP
concentration. At the most damage-intensive condition,
$L=50~\mathrm{keV\,\mu m^{-1}}$ and a $20\times$ GNP concentration,
the model predicts $R_c=73.1$, $R_m=170.8$, and
$U(24\,\mathrm{h})=178.6$ DSB lesions per nucleus, compared with the
Geant4-DNA-derived values of $36.1$, $204.7$, and $181.7$,
respectively. Since both endpoint sets conserve the same initial
burden, $N_0=422.5$ DSB lesions per nucleus, their normalized
partitions are
\begin{equation}
\left(
\frac{R_{c,\mathrm{mod}}}{N_0},
\frac{R_{m,\mathrm{mod}}}{N_0},
\frac{U_{\mathrm{mod}}}{N_0}
\right)
=
(0.173,\,0.404,\,0.423),
\end{equation}
and
\begin{equation}
\left(
\frac{R_{c,\mathrm{G4}}}{N_0},
\frac{R_{m,\mathrm{G4}}}{N_0},
\frac{U_{\mathrm{G4}}}{N_0}
\right)
=
(0.085,\,0.485,\,0.430).
\end{equation}
\begin{table*}[h]
\centering
\caption{Comparison of independently predicted repair endpoints with the
Geant4-DNA-derived endpoint data for AG1522 cells following a proton
dose of $D=2~\mathrm{Gy}$.  The notation M and G4 denotes the model
prediction and Geant4-DNA-derived value, respectively.}
\label{tab:endpoint-validation-2Gy}
\begin{tabular}{ccccccccc} \hline\hline
& & Correct Repair, $R_{c}$  &  Misrepair, $R_{m}$  &  Residual damage, $U(24,\mathrm{h})$ \\
LET &  GNP &  M/G4  &  M/G4  &  M/G4  \\
$(\mathrm{keV}\,\mu\mathrm{m}^{-1})$ & concentration & &  DSB lesions per nucleus  \\ \hline
10 & $5\times$  & 111.0/110.3 &  18.1/17.9 &   8.0/8.9  \\
10 & $10\times$ & 115.4/109.5 &  23.4/27.4 &  10.0/11.9 \\
10 & $15\times$ & 116.4/105.7 &  29.2/37.2 &  12.3/15.0 \\
10 & $20\times$ & 114.8/99.9 &  35.7/47.0 &  14.6/18.2 \\
\addlinespace

20 & $5\times$  & 133.1/128.7 &  39.7/45.2 &  21.5/ 20.4\\
20 & $10\times$ & 135.2/126.8 &  52.4/62.5 &  28.7/27.0 \\
20 & $15\times$ & 132.2/120.6 &  66.1/80.4 &  36.8/34.1 \\
20 & $20\times$ & 125.7/111.4 &  80.2/98.8 &  45.8/41.5 \\
\addlinespace

30 & $5\times$  & 130.5/119.3 &  66.3/79.5 &  37.1/35.1 \\
30 & $10\times$ & 129.5/112.6 &  84.7/104.0 &  50.0/47.6 \\
30 & $15\times$ & 123.5/101.2 & 103.0/128.9 &  64.8/61.2 \\
30 & $20\times$ & 114.7/ 86.5 & 120.2/153.9 &  81.4/75.9 \\
\addlinespace

40 & $5\times$  & 117.4/103.2 & 103.0/116.3 &  61.0/61.9 \\
40 & $10\times$ & 113.8/92.4 & 126.1/144.6 &  82.6/85.5 \\
40 & $15\times$ & 106.3/77.0 & 146.8/171.5 & 107.1/111.7 \\
40 & $20\times$ &  97.2/61.7 & 164.1/195.4 & 134.3/138.5  \\
\addlinespace

50 & $5\times$  &  91.7/76.2 & 121.0/135.5 &  81.2/82.3 \\
50 & $10\times$ &  87.3/63.3 & 142.2/162.7 & 110.2/113.8 \\
50 & $15\times$ &  80.5/49.4 & 159.0/185.8 & 142.9/147.2  \\
50 & $20\times$ &  73.1/36.1 & 170.8/204.7 & 178.6/181.7  \\ \hline
RMSE & & 20.53 & 19.81 & 3.14\\
MAE &  & 17.88 & 17.37 & 2.85\\ \hline\hline
\end{tabular}
\end{table*}

Thus, the  kinetic formulation accurately captures the
persistent lesion fraction but assigns too many processed lesions to
correct repair and too few to misrepair. At this condition, correct
repair is overpredicted by $37.0$ lesions per nucleus, misrepair is
underpredicted by $33.9$ lesions per nucleus, and residual damage
differs by only $-3.1$ lesions per nucleus. This pattern indicates
that the independently inferred complex-lesion removal coefficient
$k_C$, together with $k_{SC}=0$, provides an adequate description of
24-h lesion persistence, whereas the class-specific fidelity closure
defined by $\mu_S$ and $\mu_C$ remains too optimistic at high damage
complexity. Consequently, the residual-damage predictions provide
the strongest validation of the kinetic model, while the
correct-repair and misrepair estimates require further constraint
from independent measurements or simulations of correct-end
joining, mutation, and chromosome misrejoining.

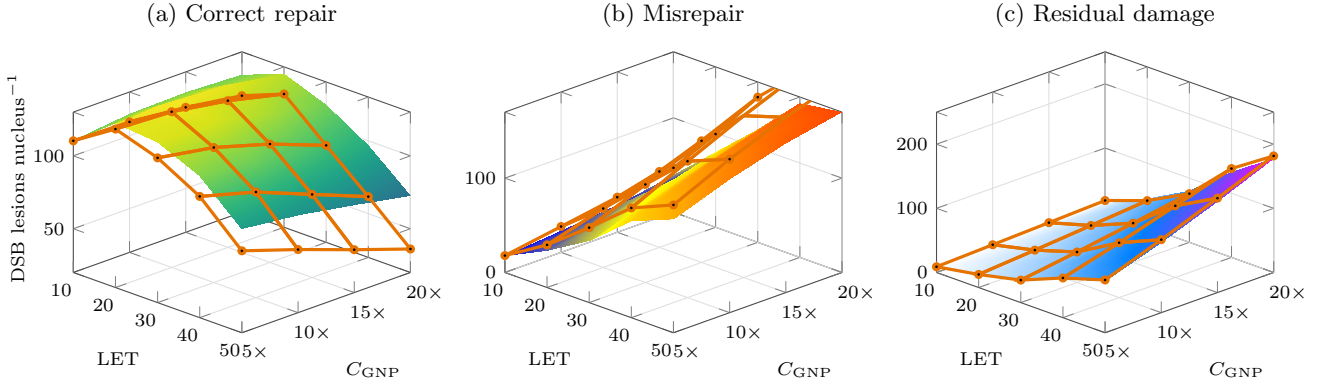
\begin{figure*}[h]
\centering

\begin{tikzpicture}
\begin{groupplot}[
    group style={
        group size=3 by 1,
        horizontal sep=1.25cm
    },
    width=0.325\textwidth,
    height=0.285\textwidth,
    view={45}{28},
    xlabel={LET},
    ylabel={$C_{\mathrm{GNP}}$},
    xmin=10, xmax=50,
    ymin=5,  ymax=20,
    xtick={10,20,30,40,50},
    ytick={5,10,15,20},
    yticklabels={$5\times$,$10\times$,$15\times$,$20\times$},
    xlabel style={font=\scriptsize},
    ylabel style={font=\scriptsize},
    zlabel style={font=\scriptsize},
    tick label style={font=\scriptsize},
    title style={font=\small},
    grid=major,
    grid style={gray!25},
    axis line style={black!70},
    scaled z ticks=false,
    zticklabel style={
        /pgf/number format/fixed,
        /pgf/number format/precision=0
    }
]

 \nextgroupplot[
    title={(a) Correct repair},
    zlabel={DSB lesions nucleus$^{-1}$},
    zlabel style={
        font=\scriptsize,
        rotate=360,
        at={(axis description cs:-0.12,0.55)},
        anchor=south
    },
    zmin=20,
    zmax=130
]


\addplot3[
    surf,
    shader=interp,
    mesh/rows=5,
    opacity=0.78,
    colormap/viridis
]
coordinates {
    (10, 5,111.0)  (10,10,115.4)  (10,15,116.4)  (10,20,114.8)
    (20, 5,133.1)  (20,10,135.2)  (20,15,132.2)  (20,20,125.7)
    (30, 5,130.5)  (30,10,129.5)  (30,15,123.5)  (30,20,114.7)
    (40, 5,117.7)  (40,10,113.8)  (40,15,106.3)  (40,20,97.2)
    (50, 5,91.7)  (50,10,87.3)  (50,15,80.5)  (50,20,73.1)
};

\addplot3[
    mesh,
    mesh/rows=5,
    draw=orange!90!black,
    very thick,
    mark=*,
    mark size=1.1pt
]
coordinates {
    (10, 5,110.3)  (10,10,109.5)  (10,15,105.7)  (10,20,99.9)
    (20, 5,128.7)  (20,10,126.8)  (20,15,120.6)  (20,20,111.4)
    (30, 5,119.3)  (30,10,112.6)  (30,15,101.2)  (30,20,86.5)
    (40, 5,103.2)  (40,10,92.4)  (40,15,77.0)  (40,20,61.7)
    (50, 5,76.2)  (50,10,63.3)  (50,15,49.4)  (50,20,36.1)
};


\nextgroupplot[
    title={(b) Misrepair},
    zmin=0,
    zmax=170
]

\addplot3[
    surf,
    shader=interp,
    mesh/rows=5,
    opacity=0.78,
    colormap/hot
]
coordinates {
    (10, 5,18.1)  (10,10,23.4)  (10,15,29.2)  (10,20,35.7)
    (20, 5,39.7)  (20,10,52.4)  (20,15,66.1)  (20,20,80.2)
    (30, 5,66.1)  (30,10,84.7)  (30,15,103.0)  (30,20,120.2)
    (40, 5,103.0)  (40,10,126.1)  (40,15,146.8)  (40,20,164.1)
    (50, 5,121.0)  (50,10,142.2)  (50,15,159.0)  (50,20,170.8)
};

\addplot3[
    mesh,
    mesh/rows=5,
    draw=orange!90!black,
    very thick,
    mark=*,
    mark size=1.1pt
]
coordinates {
    (10, 5,17.9)  (10,10,27.4)  (10,15,37.2)  (10,20,47.0)
    (20, 5,45.2)  (20,10,62.5)  (20,15,80.4)  (20,20,98.8)
    (30, 5,79.5)  (30,10,104.0)  (30,15,128.9)  (30,20,153.9)
    (40, 5,116.3)  (40,10,144.6) (40,15,171.5) (40,20,195.4)
    (50, 5,135.5) (50,10,162.7) (50,15,185.8) (50,20,204.7)
};


\nextgroupplot[
    title={(c) Residual damage},
    zmin=0,
    zmax=250
]

\addplot3[
    surf,
    shader=interp,
    mesh/rows=5,
    opacity=0.78,
    colormap/cool
]
coordinates {
    (10, 5,8.0)  (10,10,10.0)  (10,15,12.3)  (10,20,14.6)
    (20, 5,21.5)  (20,10,28.7)  (20,15,36.8)  (20,20,45.8)
    (30, 5,37.1)  (30,10,50.0)  (30,15,64.8) (30,20,81.4)
    (40, 5,61.0)  (40,10,82.6) (40,15,107.1) (40,20,134.3)
    (50, 5,81.2) (50,10,110.2) (50,15,142.9) (50,20,178.6)
};

\addplot3[
    mesh,
    mesh/rows=5,
    draw=orange!90!black,
    very thick,
    mark=*,
    mark size=1.1pt
]
coordinates {
    (10, 5, 8.9)  (10,10, 11.9)  (10,15,15.0)  (10,20,18.2)
    (20, 5,20.4)  (20,10,27.0)  (20,15,34.1)  (20,20,41.5)
    (30, 5,35.1)  (30,10,47.6)  (30,15,61.2)  (30,20,75.9)
    (40, 5,61.9)  (40,10,85.5)  (40,15,111.7)  (40,20,138.5)
    (50, 5,82.3)  (50,10,113.8)  (50,15,147.2) (50,20,181.7)
};

\end{groupplot}
\end{tikzpicture}

\caption{Three-dimensional dependence of DSB repair endpoints on proton LET 
  and relative GNP concentration ($C_{\mathrm{GNP}}$) for AG1522 cells at 
  $t = 24$~h following a proton dose of $D_p = 1.5$~Gy. Panels show 
  (a)~correct repair, (b)~misrepair, and (c)~residual damage, expressed as the 
  number of DSB lesions per nucleus. Colored surfaces represent the 
  kinetic-model predictions, while orange wireframes denote the corresponding 
  Geant4-DNA comparison values.}
\label{fig:repair_3d}
\end{figure*}

To quantify the combined influence of LET and relative GNP concentration on the biological effectiveness of proton 
therapy, we examined the three-dimensional dependence of DNA double-strand break 
(DSB) repair endpoints. Figure~\ref{fig:repair_3d} illustrates the kinetic-model 
predictions for correct repair, misrepair, and residual damage as functions of 
both LET and $C_{\mathrm{GNP}}$ at $t = 24$~h following a proton dose of 
$D_p = 1.5$~Gy. 

As shown in Fig.~\ref{fig:repair_3d}(a), the fraction of 
correctly repaired DSB lesions exhibits a non-monotonic dependence on LET, 
initially rising to a maximum near $20\text{--}30$~keV/$\mu$m before declining 
at higher LET values, a trend that is modestly enhanced by increasing GNP 
concentrations. Conversely, Fig.~\ref{fig:repair_3d}(b) reveals that misrepair 
yields increase monotonically with LET, with the steepest rise occurring beyond 
$30$~keV/$\mu$m, indicating that higher-LET protons drive a greater proportion 
of lesions toward erroneous rejoining pathways. The cumulative effect of these 
competing processes is captured in Fig.~\ref{fig:repair_3d}(c), where residual 
damage---the unrepaired DSB fraction---shows a pronounced synergistic increase 
at high LET and elevated $C_{\mathrm{GNP}}$. Notably, the orange wireframes, 
which represent the Geant4-DNA comparison values, closely track the 
kinetic-model surfaces across the parameter space, validating the model's 
predictive capability. These results suggest that the combination of high LET 
and GNP concentration amplifies residual DNA damage, a finding with direct 
implications for optimizing proton therapy treatment planning. The comparison also demonstrates that persistent damage under
high-LET, GNP-enhanced irradiation is governed primarily by the large
initial complex-lesion population and its slower resolution, rather
than by post-irradiation conversion of simple DSBs into complex
lesions.

\subsection{Global endpoint agreement}

\begin{figure*}[h]
\centering
\includegraphics[width=\textwidth]
{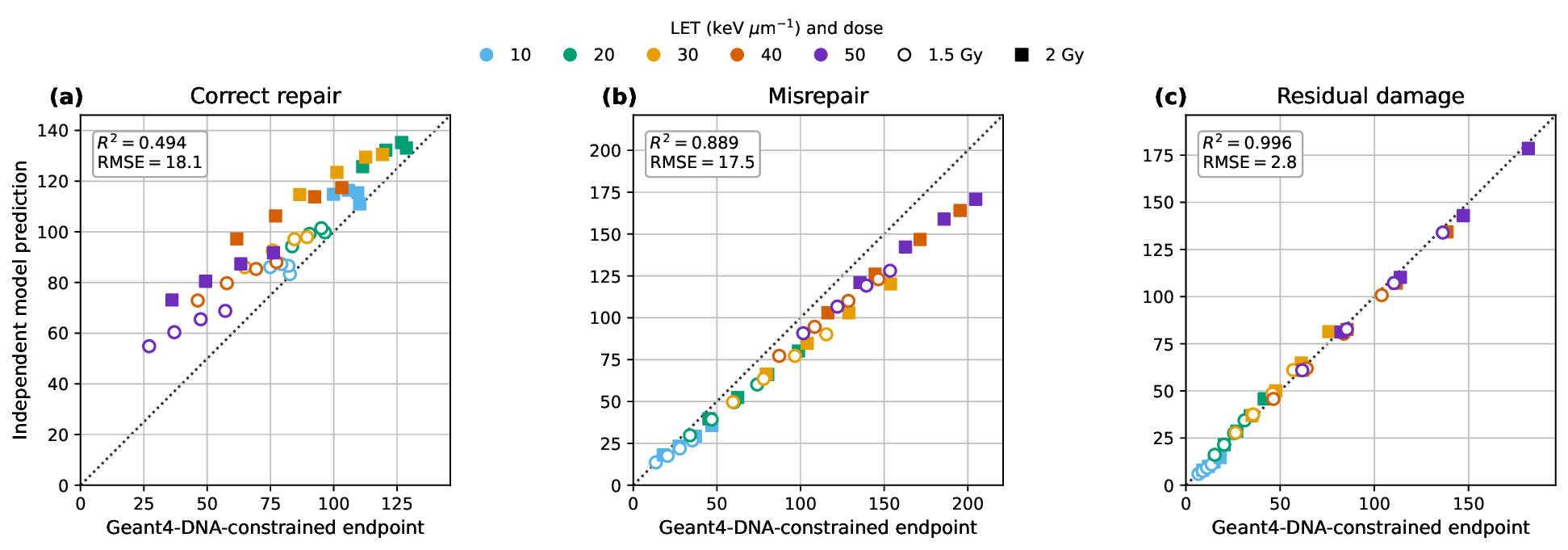}
\caption{%
Comparison of independently predicted and Geant4-DNA-constrained
repair endpoints for AG1522 cells at proton doses of $1.5$ and
$2.0$ Gy. Panels (a)--(c) show correct repair, misrepair, and
residual damage at $24$ h, respectively. Colours indicate LET in
$\mathrm{keV}\,\mu\mathrm{m}^{-1}$; open circles and filled squares
denote $1.5$ and $2.0$ Gy, respectively. The dotted line represents
perfect agreement. Predictions were obtained using leave-one-LET-out
validation.%
}
\label{fig:revised-endpoint-parity}
\end{figure*}

The endpoint comparison was extended over all 100 combinations of LET,
GNP concentration, and dose. Figure~\ref{fig:revised-endpoint-parity} compares the independently
predicted repair endpoints with the corresponding
Geant4-DNA-constrained values for AG1522 cells irradiated with proton
doses of $1.5$ and $2.0$ Gy. 

The predictions were obtained using
leave-one-LET-out validation: for each LET, all GNP concentrations and
doses associated with that LET were excluded from parameter estimation
before the endpoints were calculated. The residual-damage predictions
show excellent agreement with the Geant4-DNA results, with
$R^{2}=0.996$ and an RMSE of $2.8$ DSBs per nucleus. The residual
points remain close to the identity line over the complete range of
LET, GNP concentration, and dose, demonstrating that the
LET--concentration response surface for the complex-DSB repair
coefficient, $k_C$, captures the principal variation in the persistent
DSB burden without direct calibration to the withheld endpoints.\\

The agreement is weaker for the repair-fate endpoints. Correct repair
gives $R^{2}=0.494$ and an RMSE of $18.1$ DSBs per nucleus, while
misrepair gives $R^{2}=0.889$ and an RMSE of $17.5$ DSBs per nucleus.
The model increasingly overpredicts correct repair and underpredicts
misrepair as LET and GNP concentration increase. These opposite
deviations are most evident for the high-LET conditions and cannot be
attributed to loss of lesion balance, because the model satisfies
$N_0=R_c+R_m+U_{24}$ to numerical precision. Instead, they indicate a
systematic inconsistency between the independently specified
simple- and complex-lesion fidelities, $\mu_S$ and $\mu_C$, and the
mean-integrated fidelity used to construct the Geant4-DNA-constrained
repair endpoints. In particular, the class-specific fidelities imply a
larger effective correct-repair probability than the integrated
fidelity at high damage complexity. Accordingly, the residual-damage
panel provides the strongest independent validation of the kinetic
model, whereas the correct-repair and misrepair panels should be
interpreted as tests of the fidelity closure. Direct simulation or
experimental measurement of correct end joining, chromosomal
misrejoining, and mutation formation would be required for a fully
independent validation of these repair-fate predictions.
\subsection{Comparison with the nanoparticle-free control}
\label{sec:gnp-free-comparison}

The effects of 15-nm GNPs were evaluated relative to the
nanoparticle-free ($0\times$) AG1522-cell control at a proton dose of
$D_p=2~\mathrm{Gy}$. The enhancement of the initial DSB yield is
comparatively modest at low GNP loading but becomes substantially
greater at intermediate and high LET when the GNP concentration is
increased. A stronger nanoparticle effect is predicted for the persistent lesion
burden. The 24-h residual DSB values in the nanoparticle-free control
were 6.2, 13.8, 24.6, 43.8, and 59.6 lesions per nucleus over the same
LET range. Under the formulation, in which $k_{SC}=0$ and the
complex-lesion removal coefficient $k_C$ is inferred using
leave-one-LET-out validation, the predicted residual burdens at
$5\times$ GNP concentration are 8.0, 21.5, 37.1, 61.0, and 81.2
lesions per nucleus. These values correspond to increases of 29.0\%,
55.8\%, 50.8\%, 39.3\%, and 36.2\% relative to their respective
nanoparticle-free controls. At $20\times$, the residual burdens rise
to 14.6, 45.8, 81.4, 134.3, and 178.6 lesions per nucleus, or
approximately 2.35, 3.32, 3.31, 3.07, and 3.00 times the corresponding
control values. The enhancement of persistent damage therefore
exceeds the enhancement of the initial DSB burden, particularly at
intermediate and high LET. Within the  model, this amplification
arises from the combination of an increased initial complex-lesion
fraction and the progressive reduction of $k_C$ with LET and GNP
concentration; it does not require post-irradiation conversion of
simple DSBs into complex lesions.\\

The GNPs also alter the predicted distribution of lesions among
correct repair, misrepair, and residual damage. In the nanoparticle-free
control, the fractions of the initial DSB burden assigned to correct
repair decrease from approximately 91.4\% at
$10~\mathrm{keV\,\mu m^{-1}}$ to 50.1\% at
$50~\mathrm{keV\,\mu m^{-1}}$, while the corresponding misrepair
fractions increase from 3.3\% to 26.9\%. At $5\times$ GNP
concentration, the model predicts correct-repair fractions of 81.0\%,
68.5\%, 55.8\%, 41.7\%, and 31.2\%, together with misrepair fractions
of 13.2\%, 20.4\%, 28.3\%, 36.6\%, and 41.2\%. At $20\times$, the
correct-repair fractions decrease further to 69.5\%, 49.9\%, 36.3\%,
24.6\%, and 17.3\%, whereas the misrepair fractions are 21.6\%,
31.9\%, 38.0\%, 41.5\%, and 40.4\%. The slight reduction in the
misrepair fraction between 40 and $50~\mathrm{keV\,\mu m^{-1}}$ at
the highest GNP loading does not indicate improved repair; instead, an
increasing proportion of lesions remains unresolved at 24~h. Indeed,
the corresponding residual fraction rises from 8.8\% at
$10~\mathrm{keV\,\mu m^{-1}}$ to 42.3\% at
$50~\mathrm{keV\,\mu m^{-1}}$.\\

These results demonstrate that the predicted GNP response cannot be
represented solely by an increase in the initial number of DSBs.
Increasing LET and GNP concentration modifies both the initial lesion
spectrum and its subsequent processing, shifting the population toward
complex lesions with slower resolution and lower assumed repair
fidelity. The  model therefore identifies two distinct routes
to adverse biological outcome: incorrect processing of lesions that
leave the unrepaired compartments and prolonged persistence of lesions
that remain unresolved. Comparison with the Geant4-DNA-derived
endpoints shows that the latter mechanism is described much more
reliably. Across the conditions in Table~XIII, residual damage gives
an RMSE of 3.14 DSB lesions per nucleus, compared with 20.53 and 19.81
lesions per nucleus for correct repair and misrepair, respectively.
At high LET and GNP concentration, the model tends to overpredict
correct repair and underpredict misrepair while retaining close
agreement for the residual burden. Consequently, the residual endpoint
provides the strongest support for the $k_{SC}=0$ kinetic
description, whereas the repair-fate fractions remain
model-constrained estimates. Independent measurements of correct-end
joining, mutation, chromosome aberrations, and clonogenic survival
will be required to validate the assumed fidelity functions
$\mu_S$ and $\mu_C$ and to determine whether the predicted redistribution
between correct repair and misrepair is biologically accurate.

\section{Summary and Conclusions}

A multiscale kinetic framework has been developed to describe the
induction, recognition, processing, and biological consequences of DNA
double-strand breaks in AG1522 human fibroblasts exposed to protons in
the presence of 15-nm gold nanoparticles. The model connects proton
dose and LET, perinuclear GNP localization, local physical and chemical
enhancement, initial simple and complex DSB induction, damage
signalling, repair-pathway allocation, lesion removal, and
pathway-dependent repair fidelity.  The distinction between initial DNA damage induction, lesion-resolution
kinetics, and repair fidelity is important because an increase in initial
damage or signalling does not necessarily imply faster lesion resolution
or faithful restoration of genomic structure.\\

The primary modification presented in this study is the elimination of the
phenomenological simple-to-complex conversion term, achieved by imposing \(k_{SC}=0\).
Under this formulation, simple and complex DSBs evolve
from independently specified initial populations, and persistent
damage arises from the large initial complex-lesion burden and its
slower resolution rather than from post-irradiation reclassification
of simple lesions. A positive, dose-independent LET--GNP response
surface was used to infer the complex-lesion removal coefficient
\(k_C\). The resulting values decrease from approximately
\(0.0828~\mathrm{h^{-1}}\) at low LET and GNP loading to
\(0.0292~\mathrm{h^{-1}}\) at the highest investigated condition,
corresponding to complex-lesion half-lives of approximately
\(8.4\)--\(23.8~\mathrm{h}\). This behavior provides a physically
consistent description of the progressive persistence of clustered
damage as LET and nanoparticle concentration increase.\\

The strongest quantitative result concerns residual damage at
\(24~\mathrm{h}\). Leave-one-LET-out validation across the 1.5- and
2.0-Gy conditions produced
\(R^2=0.996\) and an RMSE of approximately 2.8 DSB lesions per nucleus.
For the 2-Gy comparison in Table~XIII, the residual endpoint gave
\(\mathrm{RMSE}=3.14\) and \(\mathrm{MAE}=2.85\) lesions per nucleus.
The close agreement obtained after withholding complete LET blocks
indicates that the  \(k_C\) response surface captures the
dominant variation in lesion persistence across the investigated LET
and GNP-concentration domain. It also supports the conclusion that the
initial complex/clustered DSB population is the principal determinant
of the subsequent residual burden.\\

Relative to the nanoparticle-free control, GNP loading increases both
the initial DSB burden and, more strongly, the persistent lesion
burden. At \(20\times\) GNP concentration and \(D_p=2~\mathrm{Gy}\),
the initial DSB yield is enhanced by approximately \(32.3\%\)--\(63.0\%\)
over the investigated LET range, whereas the 24-h residual burden is
approximately \(2.35\)--\(3.32\) times the corresponding control value.
The nanoparticle effect therefore cannot be represented adequately by
a uniform initial-dose or DSB-yield enhancement. It also changes the
lesion spectrum and the characteristic timescale of post-irradiation
processing.\\

The repair-fate predictions require a more cautious interpretation.
The model predicts that simple-DSB fidelity remains comparatively high,
decreasing from approximately \(95.5\%\) to \(91.2\%\), whereas
complex-DSB fidelity decreases from \(70.8\%\) to \(8.8\%\) as LET and
GNP concentration increase. The resulting processed-lesion-weighted
misrepair fraction rises from \(14.2\%\) to \(70.0\%\). Although these
trends are biologically plausible, agreement with the comparison
endpoints is weaker than for residual damage. Correct repair gives
\(R^2=0.494\) and an RMSE of 18.1 lesions per nucleus, while misrepair
gives \(R^2=0.889\) and an RMSE of 17.5 lesions per nucleus. At high
LET and GNP loading, the model systematically overpredicts correct
repair and underpredicts misrepair while preserving the total lesion
balance. This opposing bias identifies the class-specific fidelity
closure, rather than the residual-damage kinetics, as the principal
remaining source of model discrepancy.\\

The present results consequently provide strong internal
out-of-sample support for the kinetic description of lesion
persistence but not yet for all repair-fate or downstream biological
endpoints. The Geant4-DNA-derived comparison values are themselves
model-based and do not constitute independent experimental validation.
In addition, the perinuclear GNP distribution is idealized, the GNP
concentrations are expressed in relative units, and the fidelity
functions \(\mu_S\) and \(\mu_C\) are not yet constrained by direct
measurements of correct-end joining. Cellular heterogeneity,
time-dependent cell-cycle redistribution, chromatin-state dependence,
oxygen heterogeneity, and possible dose-rate effects are also treated
only phenomenologically or omitted. Predictions of chromosome
aberrations and clonogenic survival therefore remain conditional on
the assumed mappings from molecular repair outcomes to cellular fate.\\

The model does not yet include immune signalling, bystander effects, nanoparticle heating, replication-fork dynamics, detailed chromatin motion or organ-scale pharmacokinetics. Future work should combine independently measured repair-focus kinetics
with assays of mutation, translocation, dicentric formation, and
clonogenic survival under matched LET, dose, GNP concentration, and
intracellular-localization conditions. Explicit scoring of correct and
incorrect end joining in mechanistic repair simulations would permit
separate estimation of pathway-specific fidelities and remove the
current dependence on an integrated fidelity closure. Extension to
absolute intracellular GNP concentrations and experimentally measured
spatial distributions would further improve the physical
interpretability and transferability of the model.\\





\section*{Acknowledgement}
The author thanks H.F.  Loan and A. Bhat for valuable discussions and comments on the manuscript. 

\appendix
\section{Gamma-Poisson Distribution}
\label{sec:Gamma}
Conditioned on $\Lambda =\lambda$, the lesion count is
\begin{displaymath}
P(K=k)\mid\Lambda =\lambda ) = \frac{e^{-\lambda}}{\lambda^{k} k!}
\end{displaymath}
The unconditional probability is obtained by averaging over traversal intensities:
\begin{displaymath}
P(K=k)=\int_{0}^{\infty}P(K=k)\mid\Lambda=\lambda ) f_{\Lambda}(\lambda)d\lambda
\end{displaymath}
Substituting the Poisson and Gamma distributions gives
\begin{eqnarray}
P(K=k) &= & \int_{0}^{\infty} \frac{e^{-\lambda}}{\lambda^{k} k!} \frac{\beta^{\alpha}}{\Gamma (\alpha )} \lambda^{\alpha -1}e^{-\beta \lambda} d\lambda ,\nonumber\\
& = & \frac{\beta^{\alpha}}{k!\Gamma (\alpha)}\int_{0}^{\infty} \lambda^{k+\alpha -1}e^{-(\beta +1)}d\lambda .\nonumber
\end{eqnarray}
Using
\begin{displaymath}
\int_{0}^{b}x^{s-1}e^{-cx} dx = \frac{\Gamma (s}{c}\qquad \mathrm{with} \qquad s=k+\alpha, \qquad c=\beta +1,
\end{displaymath}
we obtain
\begin{displaymath}
P(K=k) = \frac{\beta^{\alpha}}{k!\Gamma (\alpha )} \frac{\Gamma (k+\alpha)}{(\beta +1)^{k+\alpha}}.
\end{displaymath}
Therefore,
\begin{equation}
P(K=k) = \frac{\Gamma (k+\alpha )}{k!\Gamma (\alpha )} \bigg ( \frac{\beta}{\beta +1}\bigg )^{\alpha}  \bigg (\frac{1}{\beta +1}\bigg )^{k}.
\label{EqPGD}
\end{equation}

\section{Population-mean Initail-lesionYield}
\label{app:Yield}
Let the DNA-weighted average of a quantity $f({\bf r})$ be
\begin{displaymath}
\langle f\rangle_{\mathrm{DNA}}= \frac{1}{M_{\mathrm{DNA}}}\int_{V_{n}}\rho_{\mathrm{DNA}}({\bf r})f({\bf r})d^{3}r,
\end{displaymath}
where
\begin{displaymath}
M_{\mathrm{DNA}}= \int_{V_{n}}\rho_{\mathrm{DNA}}({\bf r})f({\bf r})d^{3}r.
\end{displaymath}
Assuming that the proton dose is approximately uniform across the DNA target, $D_{p}({\bf r}) \approx D_{p}$, Eq. (\ref{eq:localdose}) beomes
\begin{equation}
\langle D_{\mathrm{loc}}\rangle_{\mathrm{DNA}} = D_{p}\left[ 1+\Bigl\langle \sum_{j=1}^{N_{NP}}\eta_{j}(L)K_{NP}( \mid {\bf r} -{\bf r}_{j} )  \Bigr\rangle_{\mathrm{DNA}} \right]. \nonumber
\end{equation}
Using Eq. (\ref{eq:eta}), we get
\begin{eqnarray}
\langle D_{\mathrm{loc}}\rangle_{\mathrm{DNA}} & = & D_{p}\left[ 1+\Bigl\langle \sum_{j}\eta_{j0}K_{NP}\Bigl\rangle_{\mathrm{DNA}}  \right. \nonumber\\
& & \left. + L \Bigl\langle \sum_{j}\eta_{j0}K_{NP}\Bigl\rangle_{\mathrm{DNA}} \right].
\end{eqnarray}
Using the perinuclear concentration field (Eq. \ref{eq:nanodistb}), and its DNA-weighted effective average $C_{pn}^{eff}= \langle C_{pn}({\bf r})\rangle_{\mathrm{DNA}}$, the two kernel sums can be reduced to
\begin{eqnarray}
\Bigl\langle \sum_{j}\eta_{j0}K_{NP}\Bigl\rangle_{\mathrm{DNA}}  & = & a_{NP}C_{pn}^{eff}, \qquad \mathrm{and}\nonumber\\
\Bigl\langle \sum_{j}\eta_{j0}K_{NP}\Bigl\rangle_{\mathrm{DNA}} & = & a_{syu}C_{pn}^{eff}.\nonumber
\end{eqnarray}
Hence,
\begin{equation}
\langle D_{\mathrm{loc}}\rangle_{\mathrm{DNA}} = D_{p}\left[ 1+a_{NP}C_{pn}^{eff}+a_{syn}C_{pn}^{eff}\right].
\label{eq:dose2}
\end{equation}
To convert dose to the expected initial lesion number, we introduce the baseline proton lesion yield $Y_{p}(L)$. More generally, 
\begin{displaymath}
N_{0}=D_{p}Y_{p}(L)+D_{p}Y_{p}(L)a_{NP}C_{pn}^{eff}+D_{p}Y_{p}(L)a_{syu}C_{pn}^{eff}L.
\end{displaymath}
Defining the effective coefficients
\begin{displaymath}
Y_{NP}(L) \equiv Y_{p}(L)a_{NP}, \qquad  Y_{syn}(L)\equiv Y_{p}(L)a_{syn},
\end{displaymath}
yields
\begin{displaymath}
N_{0}= D_{p}\left[Y_{p}(L)+Y_{NP}(L)C_{pn}^{eff}+Y_{syn}(L)C_{pn}^{eff}L \right].
\end{displaymath}
If the weak residual $L$-dependence of the effective coefficients is absorbed into fitted constants, the above equation becomes
\begin{equation}
N_{0}= D_{p}\left[Y_{p}(L)+Y_{NP}(L)C_{pn}+Y_{syn}(L)C_{pn}L \right].
\label{eq:dose2}
\end{equation}

\section{General Solutions for Correct and Misrepair Equations}
\label{sec:Solutions}
Assume
\begin{displaymath}
S(0)=S_{0}, \qquad C(0)=C_{0}, \qquad R_{c}(0)=R_{m}(0)=0,
\end{displaymath}
and define
\begin{eqnarray}
a_{S}(t) & = & g_{\mathrm{rec}}\left[P(t) \right] \sum_{j\in \{ N,H,A \}}p_{j,S}(t)k_{j,S}, \qquad \mathrm{and}\nonumber\\
a_{C}(t) & = & g_{\mathrm{rec}}\left[P(t)\right] \sum_{j\in\{ N,H,A \}}p_{j,C}(t)k_{j,C}.\nonumber
\end{eqnarray}
The DSB equations become
\begin{eqnarray}
\frac{dS}{dt} & = & -\left[ k_{SC}+a_{S}(t)\right] S \nonumber\\
\frac{dC}{dt} & = &  k_{SC}S-a_{C}(t) C.\nonumber
\end{eqnarray}
The corresponding solutions are
\begin{subequations}
\begin{align}
S(t)  =  S_{0}\mathrm{exp} \left[ -k_{SC}t - \int_{0}^{t}a_{S}(u)du\right] = S_{0}e^{\Lambda_{s}(t)},\\
C(t)  =  C_{0}e^{-\int_{0}^{t}a_{C}(u)du}+k_{SC}\int_{0}^{t}S(s)\nonumber\\
 \times \mathrm{exp}\left[ -\int_{s}^{t}a_{C}(u)du\right] ds,
\end{align}
\label{eq:sol}
\end{subequations}
where $\Lambda_{s}(t)$ in Eq. (\ref{eq:sol}a) is the cumulative simple-DSB hazard.  The first term in Eq. (\ref{eq:sol}b) represents initially complex DSBs that remain at time \(t\). The second term represents simple DSBs converted to complex intermediates at time \(s\) and surviving complex-lesion processing from \(s\) to \(t\).

The pathway fluxes are
\begin{eqnarray}
J_{j,S}(t)  & = & g_{\mathrm{rec}}\left[P(t) \right] p_{j,S}(t)k_{j,S}S(t) \nonumber\\
J_{j,C}(t)  & = & g_{\mathrm{rec}}\left[P(t) \right] p_{j,C}(t)k_{j,C}C(t). \nonumber
\end{eqnarray}
We define the class-specific correct-repair coefficients
\begin{eqnarray}
c_{S}(t)  & = & g_{\mathrm{rec}}\left[P(t) \right] \sum_{j}\mu_{j,S}p_{j,S}(t)k_{j,S} \nonumber\\
c_{C}(t)  & = & g_{\mathrm{rec}}\left[P(t) \right] \sum_{j}\mu_{j,C}p_{j,C}(t)k_{j,C}. \nonumber
\end{eqnarray}
Similarly, we define the misrepair coefficients
\begin{eqnarray}
m_{S}(t)  & = & g_{\mathrm{rec}}\left[P(t) \right] \sum_{j}(1-\mu_{j,S})p_{j,S}(t)k_{j,S} \nonumber\\
m_{C}(t)  & = & g_{\mathrm{rec}}\left[P(t) \right] \sum_{j}(1-\mu_{j,C})p_{j,C}(t)k_{j,C}. \nonumber
\end{eqnarray}
The above equations satisfy
\begin{displaymath}
c_{S}(t)+m_{S}(t)=a_{S}(t),\qquad c_{C}(t)+m_{C}(t)=a_{C}(t).
\end{displaymath}
The outcome equations reduce to
\begin{subequations}
\begin{align}
\frac{dR_{c}}{dt}  = c_{S}(t)S(t)+c_{C}(t)C(t), \\
\frac{R_{m}}{dt}  = m_{S}(t)S(t)+m_{C}(t)C(t).
\end{align}
\label{eq:repair}
\end{subequations}
The solutions of the Eqs. (\ref{eq:repair}) are
\begin{subequations}
\begin{align}
R_{c}(t)  =  \sum_{j\in \{ N,H,A \} }\int_{0}^{t} g_{\mathrm{rec}}\left[ P(s)\right]  \left[ \mu_{j,S}p_{j,S}(s)k_{j,S}S(s)\right. \nonumber ,\\
  \left. + \mu_{j,C}p_{j,C}(s)k_{j,C}C(s)\right] ds, \\
R_{m}(t)  =  \sum_{j\in \{ N,H,A\} }\int_{0}^{t} g_{\mathrm{rec}}\left[ P(s)\right]\left[(1- \mu_{j,S})p_{j,S}(s)k_{j,S}S(s)\right. \nonumber\\
 \left. + (1-\mu_{j,C})p_{j,C}(s)k_{j,C}C(s)\right] ds.
\end{align}
\label{eq:Rsol}
\end{subequations}
Equations (\ref{eq:sol}), and (\ref{eq:Rsol}) are the exact integral solutions of the time-dependent system.
For a reduced form, assume that over the relevant repair interval,
\begin{displaymath}
g_{\mathrm{rec}}[P(t)] \sim  \bar{g}_{\mathrm{rec}}, \qquad p_{j,x}\sim \bar{p}_{j,x},\qquad \mu_{j,x}=\mathrm{constant}
\end{displaymath}
and define
\begin{eqnarray}
a_{S} &=& {\bar g}_{\mathrm{rec}}\sum_{j}{\bar p}_{j,S}k_{j,S}, 
\qquad a_{C}= {\bar g}_{\mathrm{rec}}\sum_{j}{\bar p}_{j,C}k_{j,C}, \nonumber\\
\lambda_{S} & =&  k_{SC}+a_{S}, \qquad  \lambda_{C}= a_{C},\nonumber
\end{eqnarray}
the constant-coefficient analytical solutions for simple and complex DSB numbers are given by
\begin{eqnarray}
S(t) & = & S_{0}e^{-\lambda_{S}t}, \nonumber\\
C(t) & = & C_{0}e^{-\lambda_{C}t}+ \frac{k_{SC}S_{0}}{\lambda_{C}-\lambda_{S}}\bigg(e^{-\lambda_{S}t} -e^{-\lambda_{C}t} \bigg)\nonumber
\end{eqnarray}
The time-integrated simple and complex DSB populations are
\begin{eqnarray}
\mathcal{I}_{S}(t) & = & \int_{0}^{t}S(s)ds = \frac{S_{0}}{\lambda_{S}}(1-e^{-\lambda_{S}t})\\
\mathcal{I}_{C}(t) & = &  \int_{0}^{t}C(s)ds = \frac{S_{0}}{\lambda_{C}}(1-e^{-\lambda_{C}t}) \nonumber\\
& & + \frac{k_{SC}S_{0}}{\lambda_{C}-\lambda_{S}}
\left[\frac{ 1-e^{-\lambda_{S}t}}{\lambda_{S}}- \frac{ 1-e^{-\lambda_{C}t}}{\lambda_{C}}\right]
\end{eqnarray}
The corresponding forms for the correct repair and misrepair are
\begin{eqnarray}
R_{c}(t) & = & c_{S}\mathcal{I}_{S}(t)+c_{C}\mathcal{I}_{C}(t) =  \frac{c_{S}S_{0}}{\lambda_{S}}(1-e^{-\lambda_{S}t}) \nonumber\\
& & + \frac{c_{C}S_{0}}{\lambda_{C}}(1-e^{-\lambda_{C}t}) +\frac{c_{C}k_{SC}S_{0}}{\lambda_{C}-\lambda_{S}}\nonumber\\
& & \times \left[\frac{1-e^{-\lambda_{S}t}}{\lambda_{S}}- \frac{ 1-e^{-\lambda_{C}t}}{\lambda_{C}}\right],\\
R_{m}(t) & = & m_{S}\mathcal{I}_{S}(t)+m_{C}\mathcal{I}_{C}(t) =  \frac{m_{S}S_{0}}{\lambda_{S}}(1-e^{-\lambda_{S}t}) \nonumber\\
& & + \frac{m_{C}C_{0}}{\lambda_{C}}(1-e^{-\lambda_{C}t}) +\frac{m_{C}k_{SC}S_{0}}{\lambda_{C}-\lambda_{S}}\nonumber\\
& & \times \left[\frac{1-e^{-\lambda_{S}t}}{\lambda_{S}}- \frac{ 1-e^{-\lambda_{C}t}}{\lambda_{C}}\right].
\end{eqnarray}

\end{document}